\documentclass{article} 
\usepackage{arxiv, times}
\usepackage{natbib}

\usepackage{ulem}

\usepackage[utf8]{inputenc} 
\usepackage[T1]{fontenc}    
\usepackage{xurl}            
\usepackage{booktabs}       
\usepackage{amsfonts}       
\usepackage{nicefrac}       
\usepackage{microtype}      
\usepackage{xcolor}         
\usepackage{hyperref}
\usepackage{soul}
\usepackage{subfigure, graphicx, tabularx}
\usepackage{amsthm, enumitem, bm}
\usepackage{xspace, float, wrapfig, comment, enumitem}
\usepackage{amsmath, amssymb, float, csquotes, pifont}
\usepackage{caption}
\hypersetup{
    colorlinks,
    linkcolor={blue!50!black},
    citecolor={blue!50!black},
    urlcolor={blue!80!black}
}
\newcommand{\cmarkg}{\textcolor{green}{\ding{51}}} 
\newcommand{\xmarkr}{\textcolor{red}{\ding{55}}} 

\usepackage{algorithm}
\usepackage{algpseudocode}

\newcommand{\prb}[1]{\textnormal{\scshape #1}}

\DeclareMathOperator*{\argmax}{arg\,max}

\makeatletter
\newtheorem*{rep@theorem}{\rep@title}
\newcommand{\newreptheorem}[2]{%
\newenvironment{rep#1}[1]{%
 \def\rep@title{#2 \ref{##1}}%
 \begin{rep@theorem}}%
 {\end{rep@theorem}}}
\makeatother

\makeatletter
\newtheorem*{rep@corollary}{\rep@title}
\newcommand{\newrepcorollary}[2]{%
\newenvironment{rep#1}[1]{%
 \def\rep@title{#2 \ref{##1}}%
 \begin{rep@corollary}}%
 {\end{rep@corollary}}}
\makeatother

\makeatletter
\newtheorem*{rep@lemma}{\rep@title}
\newcommand{\newreplemma}[2]{%
\newenvironment{rep#1}[1]{%
 \def\rep@title{#2 \ref{##1}}%
 \begin{rep@lemma}}%
 {\end{rep@lemma}}}
\makeatother

\makeatletter
\newtheorem*{rep@proposition}{\rep@title}
\newcommand{\newrepproposition}[2]{%
\newenvironment{rep#1}[1]{%
 \def\rep@title{#2 \ref{##1}}%
 \begin{rep@proposition}}%
 {\end{rep@proposition}}}
\makeatother

\newrepproposition{proposition}{Proposition}
\newreptheorem{theorem}{Theorem}
\newrepcorollary{corollary}{Corollary}
\newreplemma{lemma}{Lemma}
\newtheorem{assumption}{Assumption}
\newtheorem{corollary}{Corollary}
\newtheorem{proposition}{Proposition}

\definecolor{lavender}{rgb}{0.90, 0.90, 0.99}
\definecolor{mint}{rgb}{0.88, 0.92, 0.88}
\definecolor{almond}{rgb}{1.00,0.95,0.81}
\definecolor{turquoise}{rgb}{0.90,0.945,0.961}
\definecolor{lightgray}{rgb}{0.886, 0.886, 0.886}

\usepackage{mdframed}

\mdfdefinestyle{theoremstyle}{
  linecolor=blue,
  linewidth=0pt,
  backgroundcolor=turquoise,
}
\mdfdefinestyle{corollarystyle}{
  linecolor=blue,
  linewidth=0pt,
  backgroundcolor=mint,
}
\mdfdefinestyle{remarkstyle}{
  linecolor=blue,
  linewidth=0pt,
  backgroundcolor=almond,
}
\mdfdefinestyle{defstyle}{
  linecolor=lightgray,
  linewidth=3pt,
  backgroundcolor=lightgray,
}

\newmdtheoremenv[style=theoremstyle, innerleftmargin =5pt, innerrightmargin =5pt, innertopmargin=1em]{theorem}{Theorem}
\newmdtheoremenv[style=corollarystyle, innerleftmargin =5pt, innerrightmargin =5pt, innertopmargin=1em]{lemma}{Lemma}
\newmdtheoremenv[style=remarkstyle, innerleftmargin =5pt, innerrightmargin =5pt, innertopmargin=1em]{remark}{Remark}
\newmdtheoremenv[style=defstyle, innerleftmargin =5pt, innerrightmargin =5pt, innertopmargin=1em]{definition}{Definition}

\title{\centering{Auction-Based Regulation for Artificial Intelligence}}

\author{%
  Marco Bornstein\thanks{Correspondence to Marco Bornstein: \texttt{marcobornsteinresearch@gmail.com}.}\thanks{{Department of Computer Science, University of Maryland, College Park, MD, USA.}} \\
  \And
  Zora Che\footnotemark[2] \\
  \And
  Suhas Julapalli\footnotemark[2] \\
  \And
  Abdirisak Mohamed\footnotemark[2]\thanks{SAP Labs, LLC.} \\
  \AND
  Amrit Singh Bedi\thanks{Department of Computer Science, University of Central Florida, FL, USA.} \\
  \And
  Furong Huang\footnotemark[2] \\
}

\begin{document}

\maketitle

\vspace{2mm}
\begin{abstract}
  In an era of \enquote{moving fast and breaking things}, regulators have moved slowly to pick up the safety, bias, and legal debris left in the wake of broken Artificial Intelligence (AI) deployment. While there is much-warranted discussion about how to address the safety, bias, and legal woes of state-of-the-art AI models, rigorous and realistic mathematical frameworks to regulate AI are lacking. Our paper addresses this challenge, proposing an auction-based regulatory mechanism that provably incentivizes agents (i) to deploy compliant models and (ii) to participate in the regulation process. We formulate AI regulation as an all-pay auction where enterprises submit models for approval. The regulator enforces compliance thresholds and further rewards models exhibiting higher compliance than their peers. We derive Nash Equilibria demonstrating that rational agents will submit models exceeding the prescribed compliance threshold. Empirical results show that our regulatory auction boosts compliance rates by $20\%$ and participation rates by $15\%$ compared to baseline regulatory mechanisms, outperforming simpler frameworks that merely impose minimum compliance standards.
 \end{abstract}

\section{Introduction}
\label{sec:intro}

The recent large-scale deployment of artificial intelligence (AI) models, such as large language models (LLMs), has simultaneously boosted human productivity while sparking concern over safety (\textit{e.g.,} hallucinations, bias, and privacy \citep{huang2025survey}).
Many industry leaders, such as Google and OpenAI, remain embroiled in controversy surrounding bias and misinformation \citep{brewster2024, robertson2024, white2024}, safety \citep{jacob2024, seetharaman2024, white2023}, as well as legality and ethics \citep{bruell2023, metz2024, moreno2023} in their development and deployment of LLMs. 
Furthermore, irresponsible LLM deployment risks the spread of misinformation or propaganda by adversaries \citep{barman2024, neumann2024diverse, sun2024exploring}. 
To date, a consistent and industry-wide solution to oversee responsible AI deployment remains elusive.

Naturally, one solution to mitigate these dangers is to increase governmental regulation over AI deployment. 
In the United States, there have been some strides, on federal \citep{whitehouse2023} and state levels \citep{cabill2024}, to regulate the safety and security of large-scale AI systems (including LLMs).
While these recent executive orders and bills highlight the necessity to develop safety standards and enact safety and security protocols, few details are offered.
This follows a consistent trend of well-deserved scrutiny towards the lack of AI regulation without providing an answer on \textit{how to develop rigorous and realistic mathematical frameworks to achieve AI regulation}. 

We believe that a rigorous and realistic mathematical framework for AI regulation consists of four key pieces: \textbf{(a)} the ability to model and to analyze participant decisions, \textbf{(b)} the existence of an \enquote{optimal} participant equilibrium, \textbf{(c)} limited mathematical assumptions, and \textbf{(d)} straightforward implementation of the framework by a regulator.
This work takes a first step towards unlocking each of these four keys, designing a regulatory framework to not only enforce strict compliance, \textit{e.g.,} safety or ethical compliance, of deployed AI models, but simultaneously to incentivize the production of more compliant AI models.
\begin{figure*}[!tbp]
    \centering
    \includegraphics[width=\linewidth]{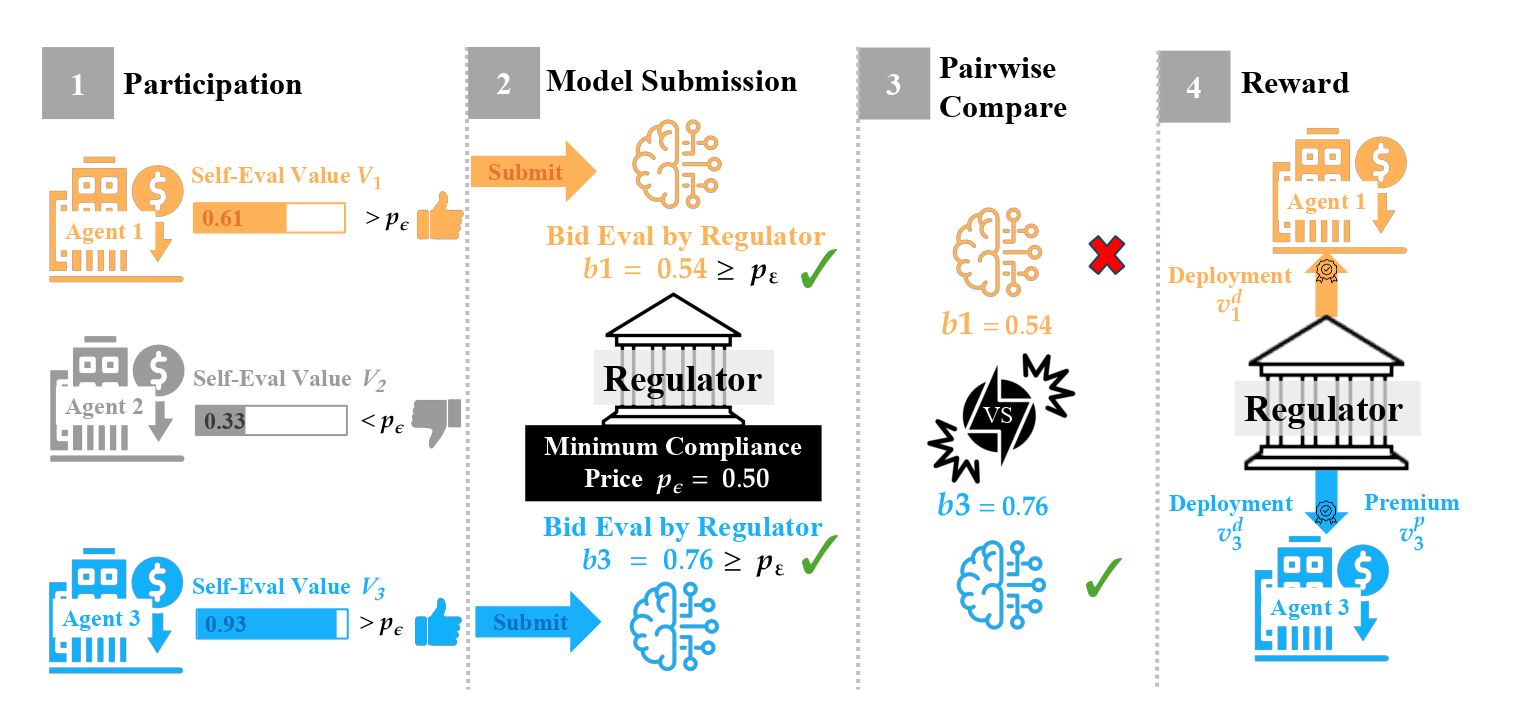}
    \caption{
    \textbf{Step-by-Step \textsc{Circa} Schematic.} 
    (Step 0) The regulator sets a compliance threshold, $\epsilon$, having corresponding price, $p_\epsilon$, required to achieve $\epsilon$.
    (Step 1) Agents evaluate their total value, $V_i$, from model deployment value ($v_i^d$) and potential regulator compensation ($v_i^p$). 
    Agents only participate if their total value exceeds $p_\epsilon$.
    (Step 2) Participating agents submit their models to the regulator, accompanied by their bid $b_i$, which reflects the amount spent to improve their model's compliance level.
    Models with bids below $p_\epsilon$ are automatically rejected.
    (Step 3) The submitted models are randomly paired, and the more compliant model (\textit{i.e.,} the higher bid) in each pair wins. 
    In this example, agent 3 wins since $b_3 > b_1$.
    (Step 4) Winning models receive both a premium and deployment value (\textit{i.e.,} agent 3 wins premium $v_3^p$ and deployment $v^d_3$ values), while losing models receive only the deployment value (\textit{i.e.,} agent 1 only wins deployment value $v^d_1$).
    }
    \label{fig:circa-schematic}
\end{figure*}

Specifically, we \textbf{(a)} formulate the AI regulatory process as an \textit{all-pay auction}, where agents (enterprises) submit their models to a regulator.
This novel auction-based regulatory mechanism leverages a reward-payment protocol that \textbf{(b)} emits Nash Equilibria at which agents \textit{deploy models that are more compliant than a prescribed threshold}. 
Analysis of our auction-based approach \textbf{(c)} requires few assumptions.
While inclusion of assumptions is non-ideal, the usage of these assumptions allows us to advance AI regulation within a sparse, yet critical, area of research.
We note, however, that the two assumptions used in this work are used within existing regulatory and AI settings \citep{goulder2013carbon, howe2024effects, rajpurkar-etal-2016-squad, stavins2008meaningful, FDA_E19_2019, N18-1101, trading_inference_time_compute} (Section \ref{sec:setting}).
Finally, our approach is \textbf{(d)} simple and can easily be implemented by a regulator (Figure \ref{fig:circa-schematic}).
Like existing regulatory frameworks \citep{coglianese2007regulation, powell2014science, van2016drugs},  we only require the regulator to: (i) prohibit deployment of models that fail to meet prescribed compliance thresholds, and (ii) incentivize compliant model production and deployment by providing additional rewards to agents that submit more compliant models than their peers.

We summarize our contributions as follows: 

\noindent \textbf{(1)} \textbf{AI Regulation:} We propose a Compliance-Incentivized Regulatory-Centered Auction (\textsc{Circa}), offering a novel approach towards AI regulation.

\noindent \textbf{(2)} \textbf{Compliance-First:} We establish, through derived Nash Equilibria, that agents are incentivized to submit models surpassing the required compliance threshold.

\noindent \textbf{(3)} \textbf{Effectiveness:} Our empirical results show that \textsc{Circa} increases model compliance by over 20\% and boosts participation rates by 15\% compared to baseline regulatory mechanisms.

\begin{table}[!t]
    \centering
    \caption{Comparison of AI regulatory frameworks across properties. \cmarkg\ denotes property is satisfied; \xmarkr\ denotes it is not.}
    \label{tab:comparison}
    \renewcommand{\arraystretch}{1.3}
    \setlength{\tabcolsep}{6pt}
    \begin{tabularx}{\textwidth}{@{} l X X X X @{}}
        \toprule
        \textbf{Feature}
        & \shortstack{\textsc{Circa} \\ \textit{(This Paper)}}
        & \citet{jagadeesan2024safety}
        & \citet{yaghini2024regulation}
        & \shortstack{All-Pay Auctions \\ \textit{(General)}} \\
        \midrule
        Overview
        & Formulates AI regulation as an auction to derive Nash Equilibria.
        & Penalizes larger companies more for unsafe AI models.
        & Introduces a multi-agent, multi-objective regulatory game.
        & Diverse formulations of all-pay auctions. \\
        \midrule
        Regulatory Scheme            & \cmarkg & \xmarkr & \cmarkg & \xmarkr \\
        Compliance-Aware             & \cmarkg & \xmarkr & \cmarkg & \xmarkr \\
        Theoretical Guarantees       & \cmarkg & \xmarkr & \xmarkr & \cmarkg \\
        Incentivizes Over-Compliance & \cmarkg & \xmarkr & \xmarkr & \xmarkr \\
        Multiple Model Builders      & \cmarkg & \cmarkg & \xmarkr & \cmarkg \\
        Single Round (Simple)        & \cmarkg & \cmarkg & \xmarkr & \cmarkg \\
        \bottomrule
    \end{tabularx}
\end{table}

\section{Related Works}\label{sec:related}
\textbf{Regulation Frameworks for Artificial Intelligence.}
A handful of work focuses on regulation frameworks for AI deployment \citep{de2021artificial, jagadeesan2024safety, rodriguez2022collusion, yew2024, qiu2025modeling, laufer2025backfiring}.
First, \citet{de2021artificial} details the need for AI regulation and surveys existing proposals. 
The proposals are ethical frameworks that express ethical decisions to make and dilemmas to address.
However, these proposals lack a mathematical framework to incentivize provably compliant models.
\citet{yew2024} provides legal analysis of copyright liability in settings with downstream, fine-tuned AI model outputs.
Existing indirect copyright liability doctrine is analyzed and deemed insufficient, with duties of care proposed as an approach for regulating AI developers.
While regulatory options are discussed, a mathematical framework with guarantees is not proposed.

\citet{rodriguez2022collusion} utilize AI models to detect collusive auctions.
This work is related to our paper but in reverse: \citet{rodriguez2022collusion} applies AI to regulate auctions and to ensure that they are not collusive.
In contrast, our paper aims to use auctions to regulate AI deployment.
\citet{laufer2025backfiring} demonstrates that regulation in a two-player AI development setting (general-purpose and domain specialist players) can backfire if not applied to both players and can improve safety if applied to both.
While equilibria are derived, our paper differs in that (i) we build a mechanism to incentivize overbidding and (ii) our setting of multiple agents and one regulator better reflects current frontier AI development. 
Both \citet{qiu2025modeling} and \citet{xu2025economics} analyze the effects of governance on model openness.
\citet{xu2025economics} considers a setting with an incumbent model developer, a downstream deployer, and an entrant developer.
Equilibrium analysis is provided, including under policy intervention, to derive the optimal level of openness for AI developers' models. 
\citet{qiu2025modeling} mathematically models how regulatory requirements on model openness affect interactions between a general-purpose model creator and a downstream specialist who fine-tunes the general-purpose model.
In our work, we construct a mechanism to incentivize overbidding and consider a competitive multi-agent, single regulator setting and not a general-specialist, two-player setting.
\citet{jagadeesan2024safety} focuses on reducing barriers to entry for smaller companies that are competing against incumbent, larger companies.
A multi-objective high-dimensional regression framework is proposed to impose \enquote{reputational damage} upon companies that deploy unsafe AI models.
Varying levels of safety constraints are allowed, and newer companies face less severe constraints in order to spur entry into the market.
This is unrealistic in practice, and only considers simple linear-regression models.

The closest related work to ours, \citet{yaghini2024regulation}, proposes a regulation game for ensuring privacy and fairness that is formulated as a Stackelberg game.
This game is a multi-agent optimization problem that is also multi-objective (for fairness and privacy). 
An equilibrium-search algorithm is presented to ensure that agents remain on the Pareto frontier of their objectives (although this frontier is estimated algorithmically).
Notably, \citet{yaghini2024regulation} considers only one model builder (agent) and multiple regulators that provide updates to the agent's strategy.
Here, a more realistic setup is considered, where there are multiple agents and a single regulator whose goal is to incentivize compliant model deployment.
It falls out of the scope of a regulator's job to collaborate with agents to optimize their strategy.
Furthermore, the mechanism proposed here is simple and efficient.
No Pareto frontier estimation or multiple rounds of optimization are required.

\textbf{All-Pay Auctions \& Contest Literature Comparison.}
Compared to the recent interest and publication of literature on regulatory frameworks for AI, all-pay auctions have been studied for decades \citep{amann1996asymmetric, baye1996all, bhaskar18, dipalantino2009crowdsourcing, gemp2022designing, goeree2000all, siegel2009all, tardos17, bertoletti2016reserve, kaplan2002all}.
These works formulate specific all-pay auctions and determine their equilibria.
Some works consider settings where agents have complete information about their rivals' bids \citep{baye1996all, bertoletti2016reserve} while others consider incomplete information, such as only knowing the distribution of agent valuations \citep{amann1996asymmetric, bhaskar18, tardos17}.

The works of \citet{amann1996asymmetric, bhaskar18, tardos17} are influential to our analysis, as their asymmetric and incomplete all-pay analysis is leveraged, in part, to help derive our Nash Equilibria.
These works, however, neither include nor analyze the effect of a reserve price (a minimum price a seller is willing to accept) on an all-pay auction.
Since our paper considers regulation, our proposed mechanism requires reserve pricing (Section \ref{sec:simple-reg}), as it is common regulatory practice to allow products that meet a compliance threshold  \citep{cfr42part84, epa2024multipollutant, epanpdes, van2016drugs}.
As a result, our paper extends the results of \citet{amann1996asymmetric, bhaskar18, tardos17} into the setting of reserve pricing.
While \citet{bertoletti2016reserve} considers an all-pay auction with a reserve price, only complete information is assumed. 
Furthermore, unlike the previous discussed works, we implement a two-tiered reward system, which involves the decomposition of agent values into deployment and premium rewards (Section \ref{sec:setting}).
This system is novel, as well as its equilibria that we derive (Corollaries \ref{cor:uniform-valuations} and \ref{cor:beta-valuations}).
This is not a trivial contribution, as the two-tiered reward system causes piecewise probability distributions that we compute and overcome during analysis (Appendix \ref{app:proofs}).
\citet{kaplan2002all} proposes an all-pay auction with variable rewards, which is related to our mechanism, but does not incorporate two-tiered rewards (it simply varies the rewards for each participating agent) and does not incorporate reserve pricing.

Furthermore, like \citet{moldovanu2006contest} and unlike the previous literature, our mechanism is not a \enquote{winner takes all} approach.
\citet{moldovanu2006contest} propose a \textit{contest architecture}, where contestants are split into several sub-contests and winners compete against one another. 
Our mechanism also allows multiple winners, where winners of the first round, those who submit compliant models, face off in a paired, two-player sub-contest with another winner.
The winner of this sub-contest wins an additional premium reward (the two-tiered reward system), which reduces the pessimism of all-pay equilibrium bids (Section \ref{sec:reg-auc}). 
Unlike \citet{moldovanu2006contest}, we propose an all-pay auction with incomplete information and reserve pricing.



\section{Regulatory Compliance of Artificial Intelligence}
\label{sec:setting}

There exists a regulator $R$ with the compliance power to set and to enforce laws and regulations (\textit{e.g.,} U.S. government regulation on lead exposure).
The regulator wants to regulate AI model deployment, by ensuring that all models meet a compliance threshold $\epsilon \in (0,1)$, \textit{e.g.,} the National Institute for Occupational Safety and Health (NIOSH) regulates that N95 respirators filter out at least 95\% of airborne particles.
If a model does not reach the compliance threshold $\epsilon$, then it is deemed unsafe and the regulator bars deployment.
On the other side, there are $n$ rational model-building agents.
Agents seek to maximize their own benefit, or utility.

\textbf{Bidding \& Evaluation}.
By law, each agent $i$ must submit, or bid in auction terminology, its model $w_i \in \mathbb{R}^d$ for evaluation to the regulator before it can be approved for deployment.
Let $S(w; x): \mathbb{R}^d \rightarrow \mathbb{R}_+$ output a compliance level (the larger the better) for model $w$ given data $x$.
In effect, each agent, given its own data $x_i$, bids a compliance level $s_i^A := S(w_i; x_i)$ to the regulator.
Subsequently, the regulator, using its own data $x_R$,  independently evaluates the agent's compliance level bid as $s_i^R := S(w_i; x_R)$.
Agent and regulator evaluation data is assumed to be independent and identically distributed (IID) $x_i, x_R \sim \mathcal{D}$.
\begin{assumption}
    \label{assumption:a1}
    Agent and regulator evaluation data comes from the same distribution $x_i, x_R \sim \mathcal{D}$.
\end{assumption}
This assumption is realistic, because agents and regulators typically rely on standardized data collection processes \citep{FDA_E19_2019} or widely accepted datasets \citep{rajpurkar-etal-2016-squad, N18-1101} for evaluation.
This ensures a fair and unbiased assessment of compliance. 
For example, FDA guidelines detail that data collection should assess efficacy and safety across various subgroups, \textit{e.g.,} demographics, while also not changing \enquote{baseline data collection determined by the clinical trial objectives} \citep{FDA_E19_2019}.
In areas such as Natural Language Processing, common datasets, or benchmarks, are employed to consistently evaluate model comprehension \citep{rajpurkar-etal-2016-squad, N18-1101}, trustworthiness \citep{wang2023decodingtrust}, and security \citep{munoz2024pyrit} across various models.
Therefore, it is reasonable to define agent $i$'s compliance level bid as $s_i := \mathbb{E}_{x \sim \mathcal{D}}[S(w_i; x)]$.
The scenario where evaluation data may be non-IID is addressed within Appendix~\ref{app:future-work}.

In regulatory settings, like the NIOSH example, a scalar compliance metric is often used.
If multiple compliance metrics must be monitored, $S$ can be defined to aggregate and weigh the various metrics.
This too is realistic in AI.
For example, LLM safety alignment literature uses a scalar-valued reward to ensure a model is aligned \citep{christiano2017deep, kaufmann2023survey, ouyang2022training}.

\textbf{Price of Compliance}.
We assume that there exists a strictly increasing function $M: (0,1) \rightarrow (0,1)$ that determines the \enquote{price of compliance} (\textit{i.e.,} maps compliance into cost).
Simply put, higher-compliant models cost more to attain.
Thus, we define the price of $\epsilon$-compliance as $p_\epsilon := M(\epsilon)$.
\begin{assumption} \label{assumption:a2}
    $\epsilon > \epsilon' \implies M(\epsilon) > M (\epsilon')$. A strictly increasing $M$ maps compliance to cost.
\end{assumption}
\vspace{-4mm}
One prominent existing example of this relationship is the cap-and-trade system that the Environmental Protection Agency exercises to combat pollution \citep{goulder2013carbon, stavins2008meaningful}.
Companies that pollute above a set emission threshold can reach compliance by purchasing allowances, or pollution deficits, from other compliant companies.
Thus, pollution compliance is attained with greater cost.
For an example within AI, models can achieve higher safety compliance through Machine Unlearning \citep{liu2024towards} or AI Alignment \citep{daisafe}.
However, such methods incur greater computational and data collection costs in exchange for improved compliance.
Furthermore, it has been found empirically that larger models and longer inference attain higher levels of compliance in adversarial training, robustness transfer, and defense \citep{howe2024effects, trading_inference_time_compute}.
However, larger models and longer inference increase training and inference costs.
We validate the compliance-cost relationship empirically in Section \ref{sec:exp}.


\textbf{Agent Costs}.
Realistically for agents, training a compliant model comes with added cost.
Consequently, each agent $i$ must decide how much money to \textit{bid}, or spend, $b_i$ to make its model compliant.
By Assumption \ref{assumption:a2}, the compliance level of an agent's model will be $s_i = M^{-1}(b_i)$.

\textbf{Agent Values}. 
\textit{(1) Model deployment value $v_i^d$.}
While it costs more for agents to produce compliant models, they gain value from having their models deployed.
Intuitively, this can be viewed as the expected value $v_i^d$ of agent $i$'s model.
The valuation for model deployment varies across agents (\textit{e.g.,} Google may value having its model deployed more than Apple).
\textit{(2) Premium reward value $v_i^p$.}
Beyond value for model deployment, the regulator can also offer additional, or premium, compensation valued as $v_i^p$ by agents (\textit{e.g.,} tax credits for electric vehicle producers or Fast Track and Priority Review of important drugs by the U.S. Food \& Drug Administration).
The regulator provides additional compensation to agents whose models exceed the prescribed compliance threshold. 
However, the value of this compensation varies across agents due to differing internal valuations. 
It is unrealistic for the regulator to compensate all agents meeting the compliance threshold due to budget constraints. 
Therefore, additional rewards are limited to a top-performing half of agents surpassing the threshold.
This ensures targeted compensation for agents enhancing compliance while maintaining feasibility for the regulator.

\textbf{Value Distribution}. 
The total value for each agent $i$ is defined as $V_i:=v_i^d + v_i^p$, which represents the sum of the deployment value and premium compensation. 
Although these values may vary widely in practice, $\{V_i\}_{i=1}^n$ is normalized for all $n$ agents to be between 0 and 1 for analytical tractability, allowing a standardized range. 
Consequently, the price to achieve the compliance threshold  $\epsilon$ is also normalized to fall within the $(0,1)$ interval, \textit{i.e.,} $p_\epsilon \in (0,1)$.
The scaling factor $\lambda_i\sim \mathcal{D}_{\lambda}(0,1/2)$ dictates the proportion of total value allocated to deployment versus compensation.
Therefore, (i) the deployment value is $v_i^d:=(1-\lambda_i)V_i$, and (ii) the premium compensation value is $v_i^p:=\lambda_iV_i$. 
Both $V_i$ and $\lambda_i$ are private to each agent, though the distributions $\mathcal{D}_V$ and $\mathcal{D}_\lambda$ are known by participants. 
The maximum allowable factor is set at $\lambda_i=1/2$, reflecting the realistic constraint that compensation should not exceed deployment value.
While Section \ref{sec:reg-auc} considers $\lambda_i\leq 1/2$, theoretical extensions can be made for scenarios where $\lambda_i > 1/2$.

\textbf{All-Pay Auction Formulation}.
Overall, agents face a trade-off: producing higher-compliant models garners value, via the regulator, but incurs greater costs.
Furthermore, in order to attain the rewards detailed above, agents must submit a model with a compliance level at least as large as $\epsilon$. 
This problem is formulated as an \textit{asymmetric all-pay auction} with \textit{incomplete information} \citep{amann1996asymmetric, bhaskar18, tardos17}.
An all-pay auction is used since agents incur an unrecoverable cost, training costs, when submitting their model to regulators.
The auction is asymmetric with incomplete information since valuations $V_i$ are private and differ for each agent.

\textbf{Auctions in Practice}. 
Above, we detail agent values as well as regulator evaluation in a theoretical manner.
In practice, before the auction begins, the regulator posts a date of deployment for all models that are submitted to the regulator and subsequently deemed \enquote{compliant} during the auction process.
Furthermore, the regulator posts an additional reward for models deemed exceedingly compliant.
As detailed above, this premium reward can include subsidies (\textit{e.g.,} tax credits of \$10,000) or fast-tracked model deployment for each agent that submits an exceedingly compliant model.
Each agent \textit{values} these rewards differently (\textit{e.g.,} art collectors, or bidders, have their own valuation for an art piece that is being auctioned off).
In the case of model deployment value $v_i^d$, Google, for example, may have internal data showcasing that having its newest Gemini model deployed will generate 3 billion dollars. 
Thus, Google would have a model deployment value of $v_i^d = 3e9$.
In summary, different agents will have different valuations for the rewards posted by the regulator.
As such, we model agent values as random variables.

Along with a date of deployment and premium reward, the regulator posts a submission deadline date.
Agents submitting models before the deadline will partake in the auction process, while agents that miss the deadline must wait for a future auction (see Appendix \ref{app:repeating-auctions-full}). 
Once the deadline passes, the regulator begins its review and evaluation of all submitted models.
Evaluation and analysis of model safety is already being done by a few institutions \citep{vanschoren2025role, bengio2025international, fli2025aisafetyindex, aisi2025frontier}.
Particularly impressive, Future of Life Institute's evaluation methodology and report \citep{fli2025aisafetyindex} is a leading and practical example of how AI models can be evaluated for compliance.
Within this approach, each agent is given a score for their submitted model.
The model is scored based on an average of expert assessments in the areas including but not limited to: (i) model performance on safety benchmarks, (ii) robustness of implemented safeguards against adversarial attacks, (iii) user privacy, (iv) watermarking and fine-tuning safeguards, (v) agents' risk identification and assessment processes, and (vi) agents' preparedness for managing extreme or existential risks.
Such a comprehensive safety evaluation is feasible for a regulator, as it is already done in practice.

\textbf{Agent Objective}.
The objective, for each model-building agent $i$, is to maximize its own utility $u_i$.
Namely, each agent seeks to determine an optimal compliance level to bid to the regulator $b_i^*$.
However, given the all-pay auction formulation, agents may need to take into account all other agents' bids $\bm{b}_{-i}$ in order to determine their optimal bid $b_i^*$,
\begin{equation}
    b_i^* := \argmax_{b} u_i(b; \; \bm{b}_{-i}).
\end{equation}
A major portion of this paper is constructing an auction-based mechanism, thereby designing the utility of each agent, such that each participating agent maximizes its utility when it bids more than \enquote{the price to obtain the compliance threshold}, \textit{i.e.,} $b_i^* > p_\epsilon$.
To begin, a simple mechanism is provided, already utilized by regulators, that does not accomplish this goal, before detailing the auction-based mechanism \textsc{Circa} that provably ensures $b_i^* > p_\epsilon$ for all agents.

\section{Reserve Thresholding: Base Regulation}
\label{sec:simple-reg}

The simplest method to ensure model compliance is for the regulators to set a reserve price, or minimum compliance level.
This mechanism is coined the \textit{multi-winner reserve thresholding auction}, where the regulator awards a deployment reward, $v_i^d$, to each agent whose model meets or exceeds the compliance threshold $\epsilon$.
Within this auction, each agent $i$'s utility is mathematically formulated as,
\begin{equation}
    \label{eq:simple-threshold}
    u_i(b; \; \bm{b}_{-i}, v_i^d) = \begin{cases}
        -b \quad &\text{ if $b < p_\epsilon$}, \\
        v_i^d - b \quad &\text{ if } b \geq p_\epsilon.
    \end{cases}
\end{equation}
However, the formulation above is ineffective at incentivizing greater than $\epsilon$-level compliance.
\begin{theorem}[Reserve Thresholding Nash Equilibrium]
\label{thm:simple-threshold}
    Under Assumption \ref{assumption:a2}, agents participating in Reserve Thresholding Equation~\ref{eq:simple-threshold} have an optimal bid and utility of,
    \begin{equation}
    \label{eq:reserve-threshold-utility}
        b_i^* = p_\epsilon, \quad u_i(b_i^*; \; \bm{b}_{-i}, v_i^d) = v_i^d - p_\epsilon,
    \end{equation}
    and submit models with the following compliance level,
    \begin{equation}
        s_i^* = \begin{cases}
            \epsilon &\text{ if } u_i(b_i^*; \; \bm{b}_{-i}, v_i^d) > 0,\\
            0 \text{ (no submission) } &\text{ else.}
        \end{cases}
    \end{equation}
\end{theorem}
\vspace{-2mm}
When a regulator implements reserve thresholding, as formally detailed in Theorem \ref{thm:simple-threshold}, agents exert minimal effort, submitting models that just meet the required compliance threshold $\epsilon$. 
While this approach ensures that all deployed models satisfy minimum compliance, it fails to encourage agents to build models with compliance levels exceeding $\epsilon$.
Additionally, agents whose deployment rewards are less than the cost of achieving compliance, \textit{i.e.,} $v_i^d < p_\epsilon$, lack incentive to participate in the regulatory process.
That lack of incentive leads to reduced participation rates (Remark~\ref{rm:lackincentive}).

\begin{remark}[Lack of Incentive]\label{rm:lackincentive}
    Each agent is only incentivized to submit a model with compliance $s_i^* = \epsilon$.
    Our goal is to incentivize agents to build models that possess compliance levels exceeding the minimum threshold: $s_i^* > \epsilon$.
\end{remark}

\section{Compliance-Incentivized Regulation: Auction-Based Approach}
\label{sec:reg-auc}

To alleviate the lack of incentives within simple regulatory auctions, such as in Section \ref{sec:simple-reg}, we propose a regulatory all-pay auction that mandates an equilibrium where agents \textit{submit models with compliance levels exceeding $\epsilon$}.

\textbf{Algorithm Description.}
The core component of the auction is that agent compliance levels are compared against one another, with the regulator rewarding those having the superior compliant model with premium compensation.
Only agents with models that achieve a compliance level of $\epsilon$ or higher are eligible to participate in the comparison process; models that do not meet this threshold are automatically rejected. 
The detailed algorithmic block of \textsc{Circa} is depicted in Algorithm~\ref{alg:circa}.

\begin{algorithm}[H]
\caption{Compliance-Incentivized Regulatory-Centered Auction (\textsc{Circa})}\label{alg:circa}
\begin{algorithmic}[1]
\State Each agent $i$ receives their total value $V_i$ and partition ratio $\lambda_i$ from \enquote{nature}
\State Agents determine their optimal bids $b_i^*$ and corresponding utility $u_i(b_i^*)$ \Comment{via Corollaries \ref{cor:uniform-valuations} or \ref{cor:beta-valuations}}
\State Agents decide to participate, the set of participating agents is $P = \{j \in [n] \; \big| \; u_j(b_j^* ; \; \textbf{b}_{-j}, v_j^d, v_j^p) > 0 \}$
\For{participating agents $j \in P$}
\State Spend $b_j^*$ to build a model, with compliance $s_j^* = M^{-1}(b_j^*)$, and submit it to the regulator
\EndFor
\State Regulator verifies compliance levels, clearing models for deployment when $s_j^* \geq \epsilon \; \forall j \in P$
\State Regulator pairs up models, awarding compensation to agents with the more compliant model
\end{algorithmic}
\end{algorithm}
\setlength{\textfloatsep}{4pt}

\textbf{Agent Utility.} The utility for each agent $i$ is therefore defined as in Equation~\ref{eq:regauc-utility-1},
\begin{equation}
    \label{eq:regauc-utility-1}
    u_i(b; \; \bm{b}_{-i}, v_i^d, v_i^p) = \big( v_i^d + v_i^p \cdot 1_{(b>b_j)} \big) \cdot 1_{(b \geq p_\epsilon)} - b.
\end{equation}
Per regulation guidelines, the compliance criteria of an accepted model must at least be $\epsilon$.
Equation~\ref{eq:regauc-utility-1} dictates that values are only realized by each agent if its model produces a bid larger than the required cost to reach $\epsilon$-compliance, $1_{(b \geq p_\epsilon)}$.
Furthermore, agents only realize additional compensation value $v_i^p$ from the regulator if their compliance level outperforms a randomly selected agent $j$, $1_{(b > b_j)}$.
Any agent that bids $b = 1$ will automatically win and realize both $v_i^d$ and $v_i^p$.
It is important to note that the cost that every agent incurs when building its model is sunk: if the model is not cleared for deployment, the cost $-b$ is still incurred. 
The agent utility is rewritten in a piece-wise manner below,
\begin{equation}
    \label{eq:all-pay-auction}
    u_i(b; \; \bm{b}_{-i}, v_i^d, v_i^p) = \begin{cases}
        - b \quad &\text{ if $b < p_\epsilon$}, \\
        v_i^d - b \quad &\text{ if $b \geq p_\epsilon$ and } b < b_j \text{ random bid } b_j,\\
        v_i^d + v_i^p - b \quad &\text{ if $b \geq p_\epsilon$ and } b > b_j.
    \end{cases}
\end{equation}
By introducing additional compensation, $v_i^p$, and, crucially, conditioning it on whether an agent's model is more compliant than that of another random agent, it becomes rational for agents to bid more than the price to obtain the minimum compliance threshold (unlike Theorem \ref{thm:simple-threshold}).

\textbf{Incentivizing Agents to Build Compliant Models.}
We establish a guarantee that agents participating in \textsc{Circa} \textit{maximize their utility with an optimal bid $b_i^*$ that is larger than \enquote{the price required to attain $\epsilon$ compliance} (i.e., $b_i^* > p_\epsilon$}) in Theorem \ref{thm:general-bidding-strategy} below.
Furthermore, agents bid in proportion to the value for additional compensation $v_i^p$ that the regulator offers for extra-compliant models.
\begin{theorem}
\label{thm:general-bidding-strategy}
    Agents participating in \prb{Circa} Equation~\ref{eq:all-pay-auction} follow an optimal bidding strategy $\hat{b}_i^*$ of,
    \begin{equation}
        \label{eq:bidding-strategy}
        \hat{b}_i^* := p_\epsilon + v_i^p F(v_i^p) - \int_0^{v_i^p} F(z)dz \; > p_\epsilon, \quad u_i(\hat{b}_i^*; \; \bm{b}_{-i}, v_i^d, v_i^p) = v_i^d - p_\epsilon + \int_0^{v_i^p} F(z)dz,
    \end{equation}
    where $F(\cdot)$ denotes the cumulative distribution function (CDF) of the random premium reward variable corresponding to the premium reward $v_i^p = V_i\lambda_i$ with $F(v_i^p) > 0$ for $v_i^p>0$.
\end{theorem}

From the result of Theorem~\ref{thm:general-bidding-strategy} we find that each participating agent within \textsc{Circa} is expected to bid more than the price of compliance $p_\epsilon$, thereby submitting a more compliant model than the required threshold.

\begin{proposition}[Expected Participating Agent Bid]
\label{prop:expected-bid}
Let $\hat{b}_i^*$ denote the optimal bidding strategy from Theorem~\ref{thm:general-bidding-strategy}. The expected participating agent's bid over the distribution of premium rewards $v_i^p$, where $F(\cdot)$ and $f(\cdot)$ denote the CDF and probability density function (PDF) of $v_i^p$, is,
\begin{equation}
    \label{eq:expected-bid}
    \mathbb{E}[\hat{b}_i^*] = p_\epsilon + \int_0^{1/2} zf(z)(1 - F(z))dz \; > p_\epsilon.
\end{equation}
\end{proposition}

\begin{remark}[Outbidding Reserve Thresholding]
    In Theorem~\ref{thm:general-bidding-strategy} and Proposition~\ref{prop:expected-bid}, participating agents submit more compliant models than the regulator requires, $s_i^* = M^{-1}\big(b^*_i\big) > \epsilon$. This improves on Theorem \ref{thm:simple-threshold}.
\end{remark}
\vspace{-2mm}

\textbf{Randomized and Deterministic Compliance Comparison.}
The standard version of \textsc{Circa} randomly pairs agents' models up against one another in a one-shot, zero-sum game.
However, more deterministic pairing methods exist and are implementable within \textsc{Circa}.
Such methods reduce the likelihood of unfair outcomes (\textit{e.g.,} a scenario where the most compliant and second-most compliant models face off).
The first alternative method is simple: repeated randomization.
In this method, the randomization process is repeated $r$ times and the number of \enquote{wins} for each agent $w_i$ is stored by the regulator.
Then, the regulator provides each agent $i$ with a fraction of the premium reward: $\frac{w_i}{r} \cdot v_i^p$.
In the case of an odd number of agents, each agent is held out $h$ times (where $h = \lceil \frac{r}{n} \rceil$) while the remaining $n-1$ agents are paired up.
The second alternative is deterministic.
In this second method, the regulator computes the empirical CDF $\hat{F}(\cdot)$ across all $n$ participating agents' model compliance bids and passes out $\hat{F}(b_i) \cdot v_i^p$ premium rewards to each agent $i$.
For example, the participating agent with the median bid would receive half of the premium reward (\textit{e.g.,} half of the regulator's proposed tax credit reward). 

\begin{remark}[Generalizable]
    Theorem \ref{thm:general-bidding-strategy} applies to any distribution for $V_i$ and $\lambda_i$ on $[0,1]$ and $[0,1/2]$, \textit{i.e.,} $V_i \sim \mathcal{D}_V(0,1)$ and $\lambda_i \sim \mathcal{D}_\lambda(0,1/2)$, respectively.
    Determining specific optimal bids, utility, and model compliance levels requires given distributions for $V_i$ and $\lambda_i$.
\end{remark}
\vspace{-2mm}

\textbf{(Special Case 1) Uniform $V_i$ and $\lambda_i$: Optimal Agent Strategy.}
Analysis of all-pay auctions \citep{amann1996asymmetric, bhaskar18, tardos17, krishna2009auction}, as well as other types of auctions, often assume a Uniform distribution for valuations.
Therefore, our first analysis of \textsc{Circa}, below in Corollary \ref{cor:uniform-valuations}, presumes Uniform distributions for $V_i$ and $\lambda_i$.
Corollary \ref{cor:uniform-valuations} determines that a participating agent's optimal strategy to maximize its utility is to submit a model with compliance levels larger than $\epsilon$ when their values $V_i$ and $\lambda_i$ come from a Uniform distribution.

\begin{corollary}[Uniform Nash Bidding]
\label{cor:uniform-valuations}
    Under Assumption \ref{assumption:a2}, for agents having total value $V_i$ and scaling factor $\lambda_i$ both stemming from a Uniform distribution, with $v_i^d = (1-\lambda_i)V_i,$ and $v_i^p = \lambda_iV_i$, their optimal bid ($b_i^* := \min \{ \hat{b}_i^*, 1\}$) and utility participating in \prb{Circa} (Equation~\ref{eq:all-pay-auction}) are,
    
    \begin{equation}
    \label{eq:uniform-optimal-bid-utility}
        \hat{b}_i^* = \begin{cases}
            p_\epsilon + \frac{(v_i^p)^2\ln(p_\epsilon)}{p_\epsilon-1}\\
            p_\epsilon + \frac{8(v_i^p)^2(\ln(2v_i^p) - 1/2)+ p_\epsilon^2}{8(p_\epsilon-1)},
        \end{cases}
        \; \; u_i(b_i^*; \; \bm{b}_{-i}, v_i^d, v_i^p) = 
        \begin{cases}
        \frac{2(v_i^p)^2\ln(p_\epsilon)}{p_\epsilon-1} + v_i^d -  b_i^* &\quad\text{if } 0 \leq v_i^p \leq \frac{p_\epsilon}{2}, \\
        \frac{2(v_i^p)^2(\ln(2v_i^p)-1) + p_\epsilon}{p_\epsilon-1} + v_i^d -  b_i^* &\quad\text{if } \frac{p_\epsilon}{2} \leq v_i^p \leq \frac{1}{2}.
        \end{cases}
    \end{equation}
    Participating agents submit models with compliance,
    \begin{equation}
    \label{eq:uniform-safety}
        s_i^* := \begin{cases}
             M^{-1}(b_i^*) > \epsilon &\text{ if } u_i(b_i^*; \; \bm{b}_{-i}, v_i^d, v_i^p) > 0,\\
            0 \text{ (no submission) } &\text{ else}.
        \end{cases}
    \end{equation}
\end{corollary}

\textbf{(Special Case 2): Beta $V_i$ and Uniform $\lambda_i$: Optimal Agent Strategy.}
In many instances, a realistic distribution for $V_i$ is a Beta distribution with $\alpha,\beta=2$.
This distribution is Gaussian-like, with the bulk of the probability density placed in the middle. 
As such, it is realistic when agent values do not congregate amongst one another and outliers (near 0 or 1) are rare.
The performance of \textsc{Circa} in this setting is analyzed in Corollary \ref{cor:beta-valuations}.
Corollary \ref{cor:beta-valuations} states that, under a Beta(2,2) distribution for $V_i$, agent $i$ maximizes its utility with an optimal bid $b_i^*$ larger than the price of $\epsilon$ compliance, $b_i^* > p_\epsilon$, resulting in a model above the $\epsilon$-compliance threshold. 
Furthermore, Corollaries \ref{cor:uniform-valuations} and \ref{cor:beta-valuations} surpass the baseline optimal bid $b_i^* = p_\epsilon$ for Reserve Thresholding (Theorem \ref{thm:simple-threshold}).

\begin{corollary}[Beta Nash Bidding]
\label{cor:beta-valuations}
    Under Assumption \ref{assumption:a2}, let agents have total value $V_i$ and scaling factor $\lambda_i$ stem from Beta ($\alpha, \beta = 2$) and Uniform distributions respectively, with $v_i^d = (1-\lambda_i)V_i$ and $v_i^p = \lambda_iV_i$.
    Denote the CDF of the Beta distribution on $[0,1]$ as $F_\beta(x) = 3x^2 - 2x^3$.
    Optimal bid and utility for agents participating in \prb{Circa} (Equation~\ref{eq:all-pay-auction}) are,
    \begin{equation}
    \label{eq:beta-optimal-bid}
    b_i^* := \min \{ \hat{b}_i^*, 1\}, \quad
        \hat{b}_i^* = \begin{cases}
        p_\epsilon + \frac{3(v_i^p)^2(p_\epsilon^2 - 2p_\epsilon + 1)}{1 - F_\beta(p_\epsilon)}  & 0 \leq v_i^p \leq \frac{p_\epsilon}{2},\\
            p_\epsilon + \frac{8(v_i^p)^2\big(6(v_i^p)^2 - 8v_i^p + 3 \big) + p_\epsilon^3(3p_\epsilon - 4)}{8(1 - F_\beta(p_\epsilon))} & \frac{p_\epsilon}{2} \leq v_i^p \leq 1/2,
        \end{cases}
    \end{equation}
    \begin{equation}
    \label{eq:beta-utility}
        u(b_i^*; \; \bm{b}_{-i}, v_i^d, v_i^p) = \begin{cases}
            v_i^d +  \frac{6(v_i^p)^2(p_\epsilon^2 - 2p_\epsilon + 1)}{1 - F_\beta(p_\epsilon)} - b_i^* \; & 0 \leq v_i^p \leq \frac{p_\epsilon}{2},\\
            v_i^d + \frac{v_i^p\big(8(v_i^p)^3 - 12(v_i^p)^2 + 6v_i^p + p_\epsilon^2(2p_\epsilon - 3)\big)}{1 - F_\beta(p_\epsilon)} - b_i^* \; & \frac{p_\epsilon}{2} \leq v_i^p \leq 1/2.
            \end{cases}
    \end{equation}
    Participating agents submit models with compliance,
    \begin{equation}
        \label{eq:beta-safety}
        s_i^* = \begin{cases}
            M^{-1}(b_i^*) > \epsilon &\text{ if } u_i(b_i^*; \; \bm{b}_{-i}, v_i^d, v_i^p) > 0,\\
            0 \text{ (no submission) } &\text{ else}.
        \end{cases}
    \end{equation}
\end{corollary}
\begin{remark}[Improved Utility \& Participation]
    \label{remark:utility}
    Through introduction of premium compensation, agent utility is improved, in Equations~\ref{eq:uniform-optimal-bid-utility} and \ref{eq:beta-utility}, versus Reserve Thresholding in Equation~\ref{eq:reserve-threshold-utility}.
    As a result, more agents break the zero-utility barrier of entry for participation, boosting both overall agent utility and participation rate.
\end{remark}
\vspace{-2mm}
The proofs of Theorems \ref{thm:simple-threshold} and \ref{thm:general-bidding-strategy}, Proposition \ref{prop:expected-bid}, as well as Corollaries \ref{cor:uniform-valuations} and \ref{cor:beta-valuations} are found within Appendix \ref{app:proofs}.
Since the premium compensation value $v_i^p$ is a product of two random variables, the PDF and CDF of $v_i^p$ becomes a piece-wise function (as shown within Appendix \ref{app:proofs}).
As a result, the optimal bidding and subsequent utility also becomes piece-wise in both Corollaries \ref{cor:uniform-valuations} and \ref{cor:beta-valuations}.
Empirically, the correctness of the computed PDF and CDFs are verified in Appendix \ref{app:experiments}.

\textbf{Why \textsc{Circa} for AI Regulation?}
We believe that \textsc{Circa} is designed for practical AI regulatory settings:

\begin{itemize}
    \item \textbf{Multiple AI Developers, One Regulator}. The current AI landscape features many simultaneously competing model builders (\textit{e.g.,} Google, OpenAI, or Mistral) with private valuations for their model's deployment (\textit{i.e.,} each generating different revenue). Furthermore, a single regulatory body is often in charge of regulating model deployment. The regulator's primary role is to evaluate submitted models and not to collaborate on their development. \textsc{Circa} is designed with this exact setting in mind.
    \item \textbf{Sunk AI Compliance Costs}.  Compliance costs in AI are sunk and non-recoverable. Agents that invest in safety alignment, adversarial robustness, or data curation cannot get their money back whether or not the resulting model gets approved. The backbone of \textsc{Circa} is an all-pay auction which models this setting.
    \item \textbf{Required AI Model Compliance}. Existing regulatory frameworks \citep{cfr42part84, epa2024multipollutant, epanpdes, van2016drugs}, only allow products to enter the market once compliance is met. With AI being deployed in increasingly high-stakes settings, this same principle must hold for AI models. AI institutions are already grading AI model compliance \citep{vanschoren2025role, bengio2025international, fli2025aisafetyindex, aisi2025frontier}, and so too can a regulator. Thus, it is necessary for an AI regulatory framework to evaluate model compliance and deploy models only once they reach a compliance threshold. \textsc{Circa} incorporates such a reserve threshold to ensure only compliant models can enter the market.
    \item \textbf{AI Over-Compliance \& Participation}. Incentivizing over-compliance is critical within AI regulatory frameworks to safeguard our society against rapid AI development. Minimum compliance thresholds today may be insufficient tomorrow due to the rapid capability growth of AI models. \textsc{Circa} smartly pits agents against one another in sub-contests to spur agents to overbid on model compliance and participate at a higher rate (Theorem \ref{thm:general-bidding-strategy} and Remark \ref{remark:utility}).
\end{itemize}

\section{Experiments}
\label{sec:exp}

We demonstrate, in Section \ref{sec:reg-auc}, that \textsc{Circa} incentivizes compliance-exceeding models and participation at higher rates than the Reserve Thresholding mechanism (Section \ref{sec:simple-reg}).
Below, we validate these theoretical results empirically.

\textbf{Lack of Baseline Regulatory Mechanisms}. 
To the best of our knowledge, there are no other comparable compliance mechanisms to regulate AI.
As a result, the Reserve Threshold mechanism that is proposed in Section \ref{sec:simple-reg} is used as a baseline.
While simple, the Reserve Threshold mechanism is a realistic baseline to compare against.
Existing regulatory bodies, like the Environmental Protection Agency (EPA), follow similar steps before clearing products (\textit{e.g.,} the EPA authorizes permits for discharging pollutants into water sources once water quality criteria are met).

\begin{figure*}[!tbp]
    \centering
    \subfigure[Uniform, $p_\epsilon = 0.25$]{
        \includegraphics[width=0.315\textwidth]{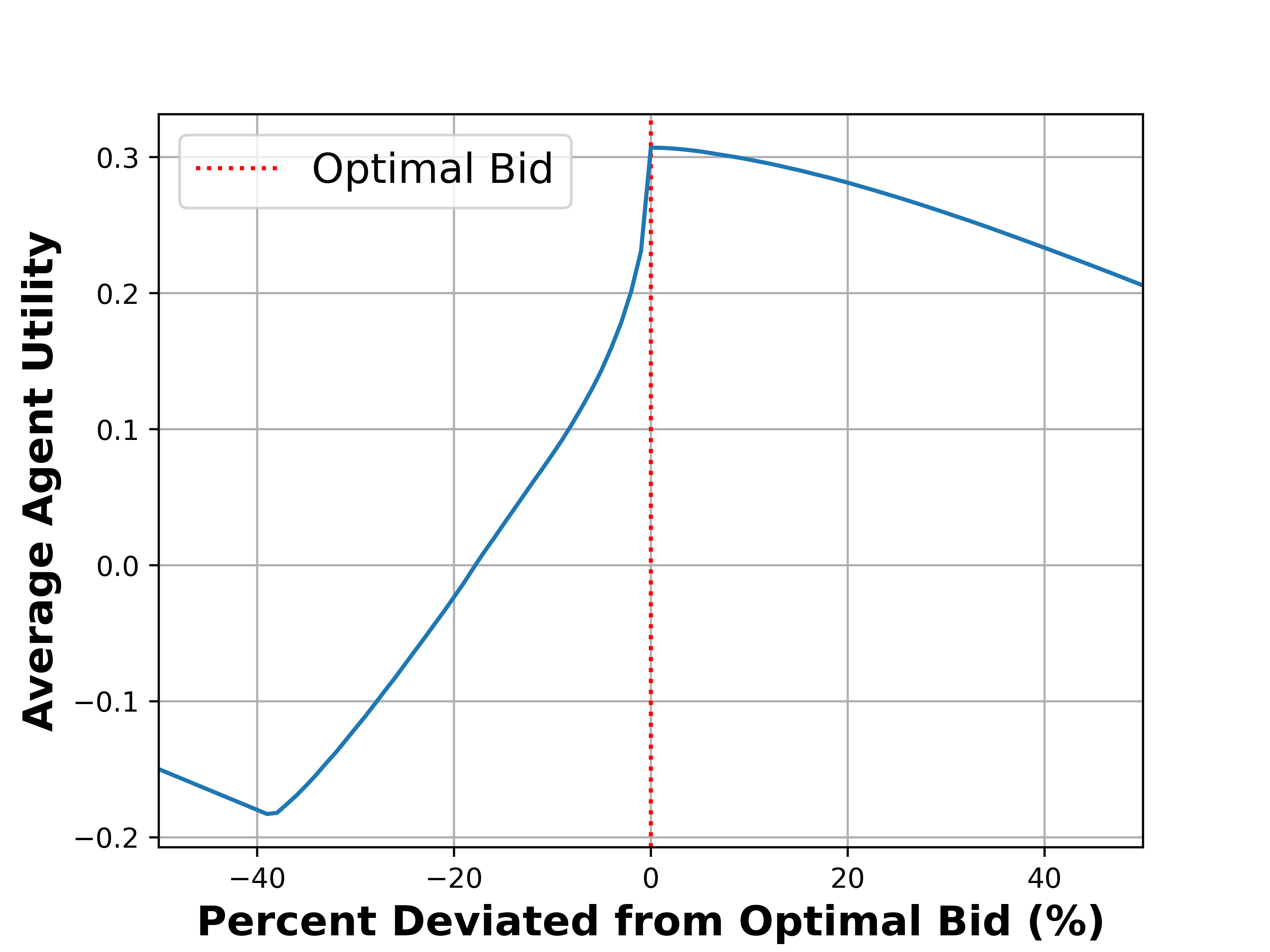}
    }
    \hfill
    \subfigure[Uniform, $p_\epsilon = 0.5$]{
        \includegraphics[width=0.315\textwidth]{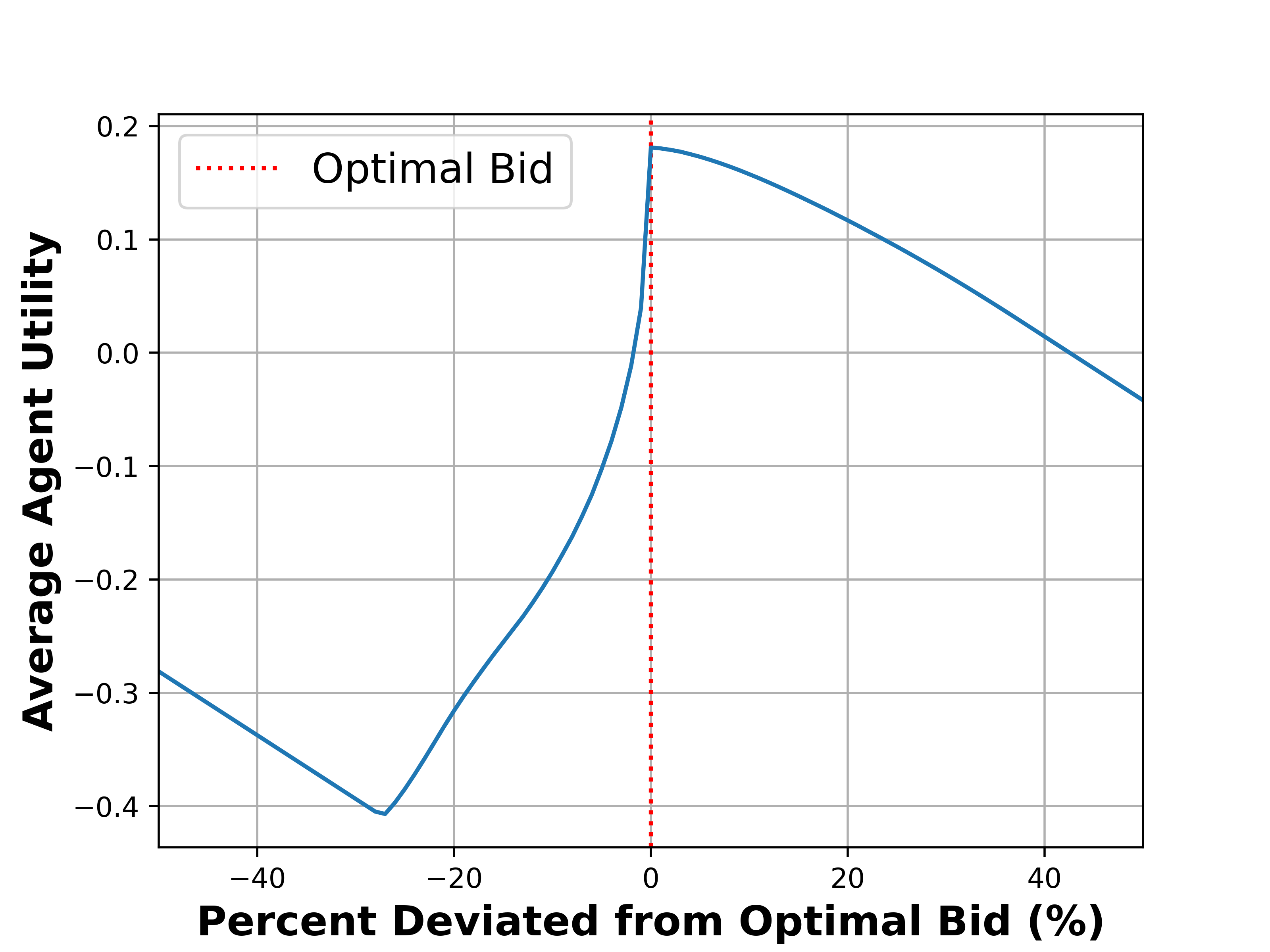}
    }
    \hfill
    \subfigure[Uniform, $p_\epsilon = 0.75$]{
        \includegraphics[width=0.315\textwidth]{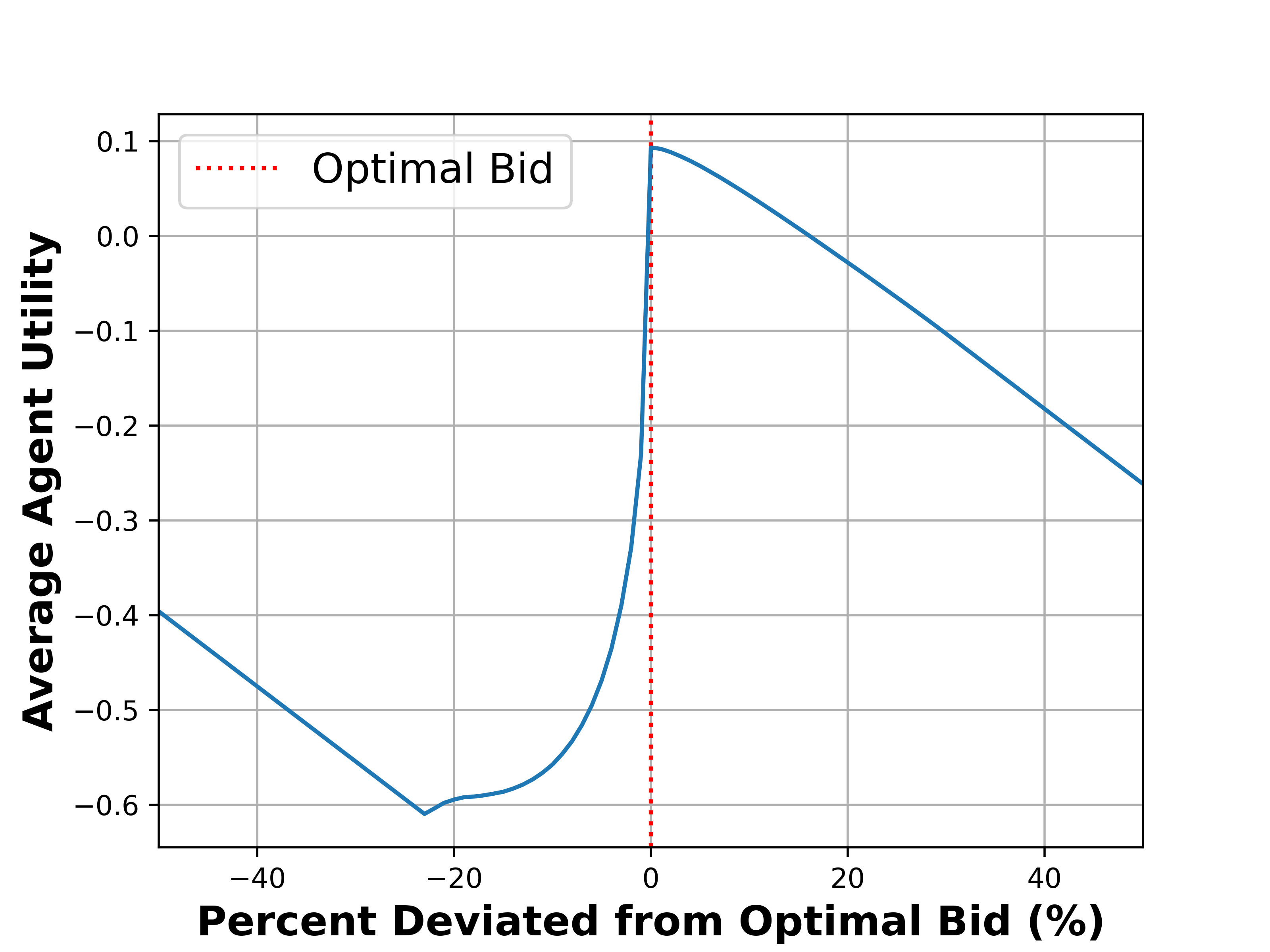}
    }

    \vspace{-3mm}

    \subfigure[Beta(2,2), $p_\epsilon = 0.25$]{
        \includegraphics[width=0.315\textwidth]{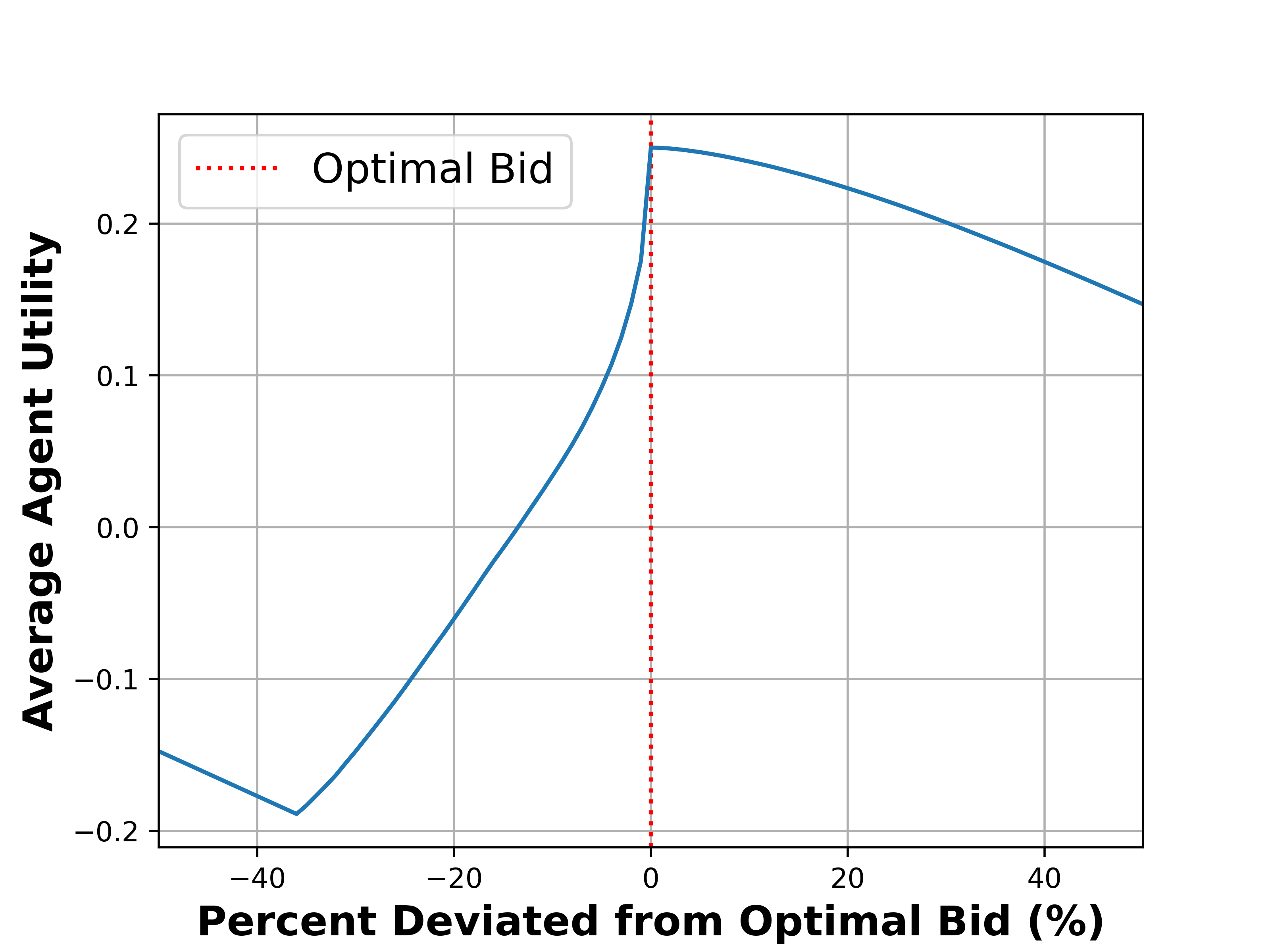}
    }
    \hfill
    \subfigure[Beta(2,2), $p_\epsilon = 0.5$]{
        \includegraphics[width=0.315\textwidth]{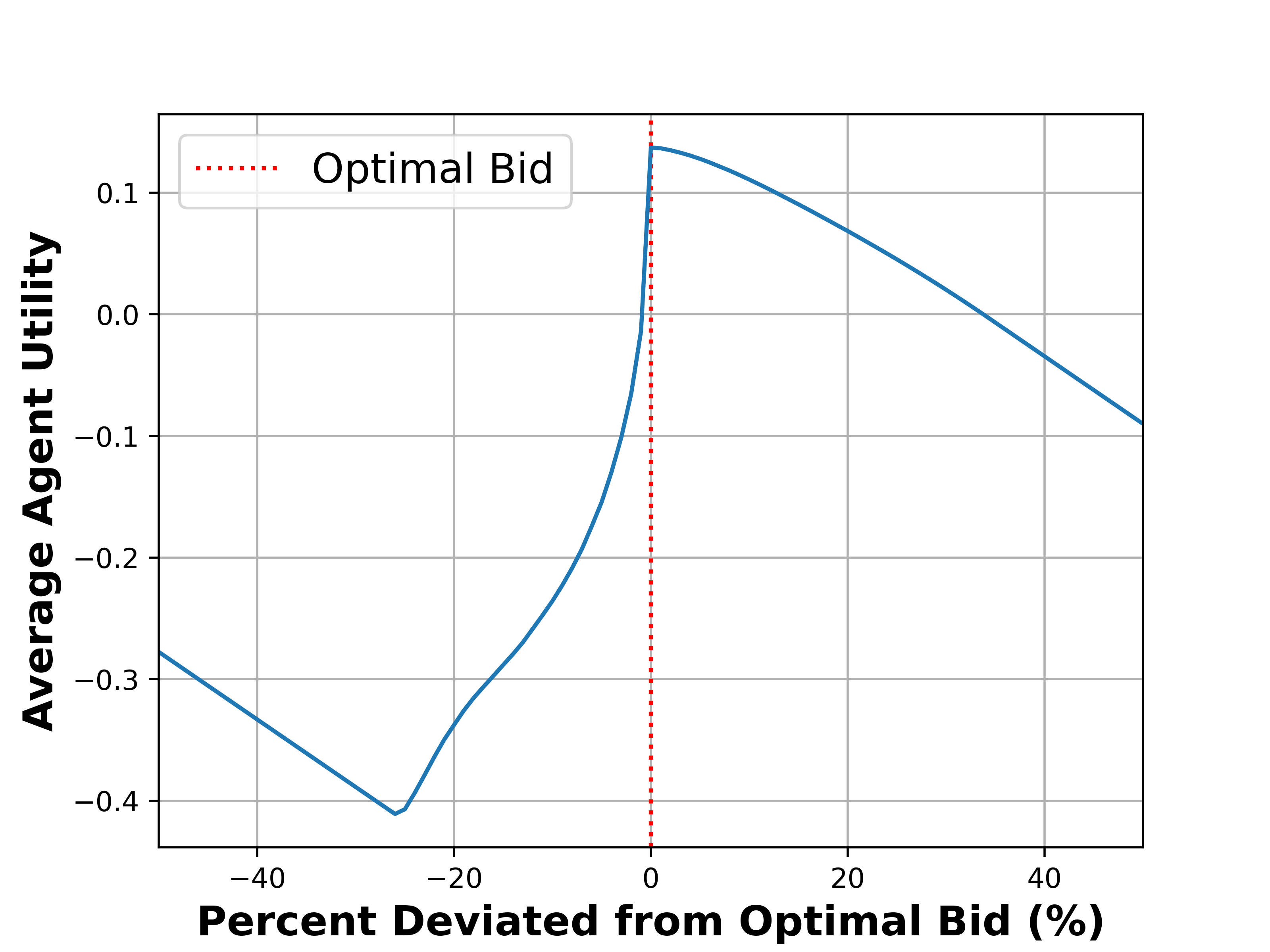}
    }
    \hfill
    \subfigure[Beta(2,2), $p_\epsilon = 0.75$]{
        \includegraphics[width=0.315\textwidth]{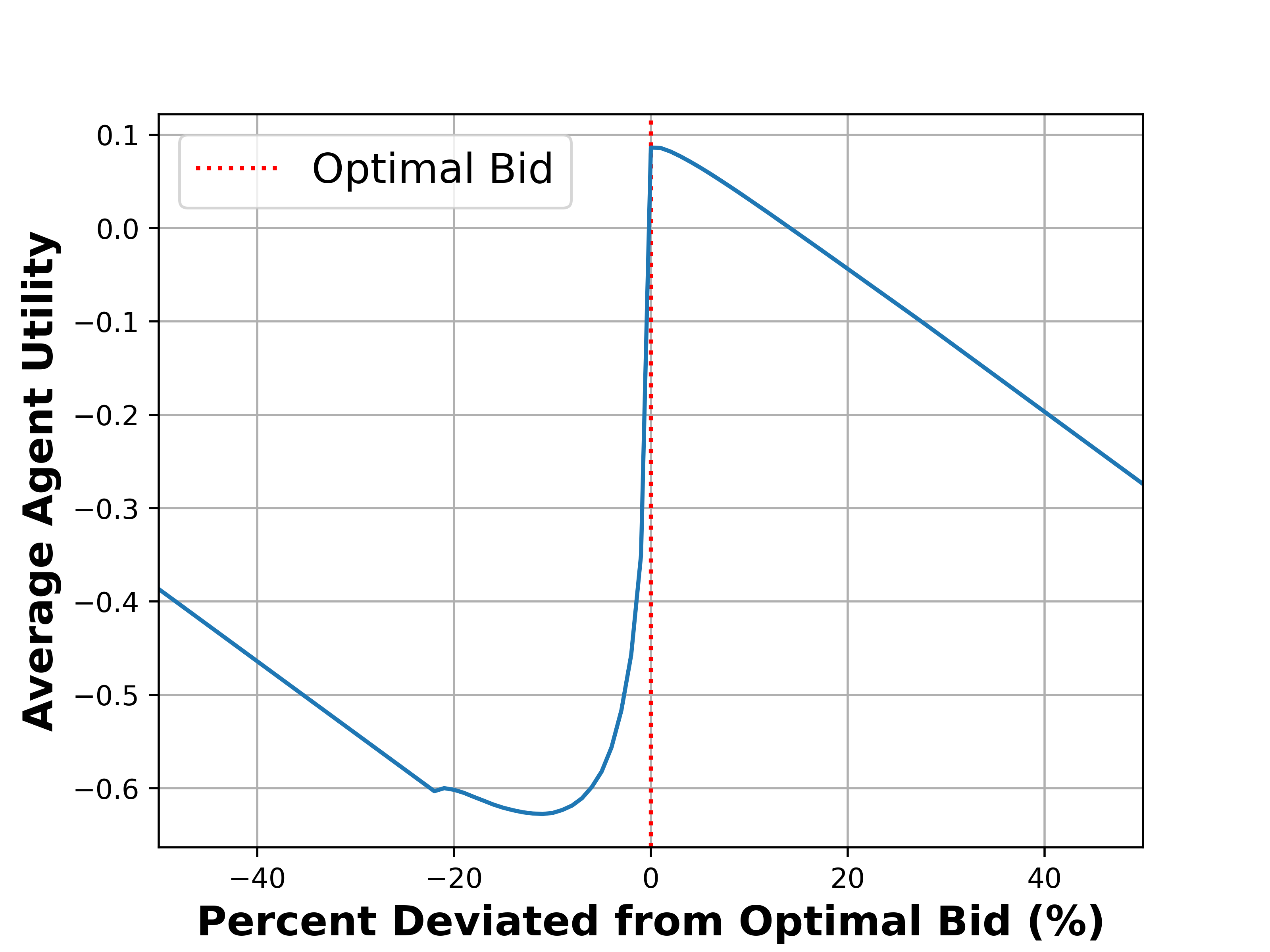}
    }

    \caption{\textbf{Validation of Nash Bidding Equilibrium.}
    Agent utility is maximized when agents follow the theoretically optimal bidding function (red line) across different distributions and varying compliance
    prices $p_\epsilon \in \{0.25, 0.5, 0.75\}$.
    \textbf{Top row:} Uniform distribution results with the optimal bid derived in Corollary~\ref{cor:uniform-valuations}.
    \textbf{Bottom row:} Beta($\alpha=\beta=2$) distribution results with the optimal bid derived in Corollary~\ref{cor:beta-valuations}. \textit{In all cases, agents attain strictly less utility when deviating from the optimal bid.}}
    \label{fig:equilibrium}
\end{figure*}

\textbf{Verifiable Nash Bidding Equilibria}.
In our first experiment, our goal is to validate that the theoretical bidding functions found in Corollaries \ref{cor:uniform-valuations} and \ref{cor:beta-valuations} constitute Nash Equilibria.
That is, an agent receives worse utility if it deviates from this bidding strategy while other agents abide by it.
To test this, we conduct a Monte Carlo simulation with $N = 100,000$ trials. 
In each trial, two agents, $i$ and $j$, independently draw their valuations from either a Uniform or Beta(2,2) distribution (Figure \ref{fig:equilibrium}).
Each agent also receives a scaling factor $\lambda_i$ that splits the total value into deployment $v_i^d = (1-\lambda_i)V_i$ and premium compensation $v_i^p = \lambda_i V_i$ values.
Once private values are provided, agents bid according to their optimal strategies for Uniform and Beta(2,2) distributions in Corollaries \ref{cor:uniform-valuations} and \ref{cor:beta-valuations} respectively.
Finally, we perturb agent $i$'s optimal bid by $\pm 50\%$ and record the utility gained or lost by participating in \textsc{Circa} across the range of bids $b_i \in [b_i^*/2, \; 3b_i^*/2]$.
We note that comparisons only occur if the other agent's bid is at least $p_\epsilon$, in order to accurately reflect how the auction mechanism in Algorithm \ref{alg:circa} functions.

When inspecting Figure \ref{fig:equilibrium}, it is clear that agent $i$'s utility is maximized at its optimal bid $b_i^*$.
Deviating from $b_i^*$ results in less utility gained for agent $i$.
Thus, \textit{our Monte Carlo simulations validate our Nash Equilibria claims}: the Uniform and Beta optimal-bidding functions in Corollaries \ref{cor:uniform-valuations} and \ref{cor:beta-valuations} indeed maximize agent utility in Figure \ref{fig:equilibrium}.
Agent $i$'s utility decays much quicker when bidding below its optimal bid $b_i^*$, since agents are \textbf{(i)} less likely to win the premium reward and \textbf{(ii)} at risk of losing the value from deployment if the bid does not reach $p_\epsilon$.
At a certain point, utility increases linearly once the agent continuously fails to bid $p_\epsilon$.
The linear improvement stems from the agent saving the cost of its bid, $-b_i$, shown in Equation~\ref{eq:all-pay-auction}.

\textbf{Improved Agent Participation and Bid Size}.
In our second experiment, we showcase that agents participate at a higher rate and have a larger expected bid when participating in \textsc{Circa} compared to our Reserve Thresholding baseline, irrespective of the compliance threshold.

\begin{figure*}[!tbp]
    \centering
    \subfigure{
    \includegraphics[width=0.44\textwidth]{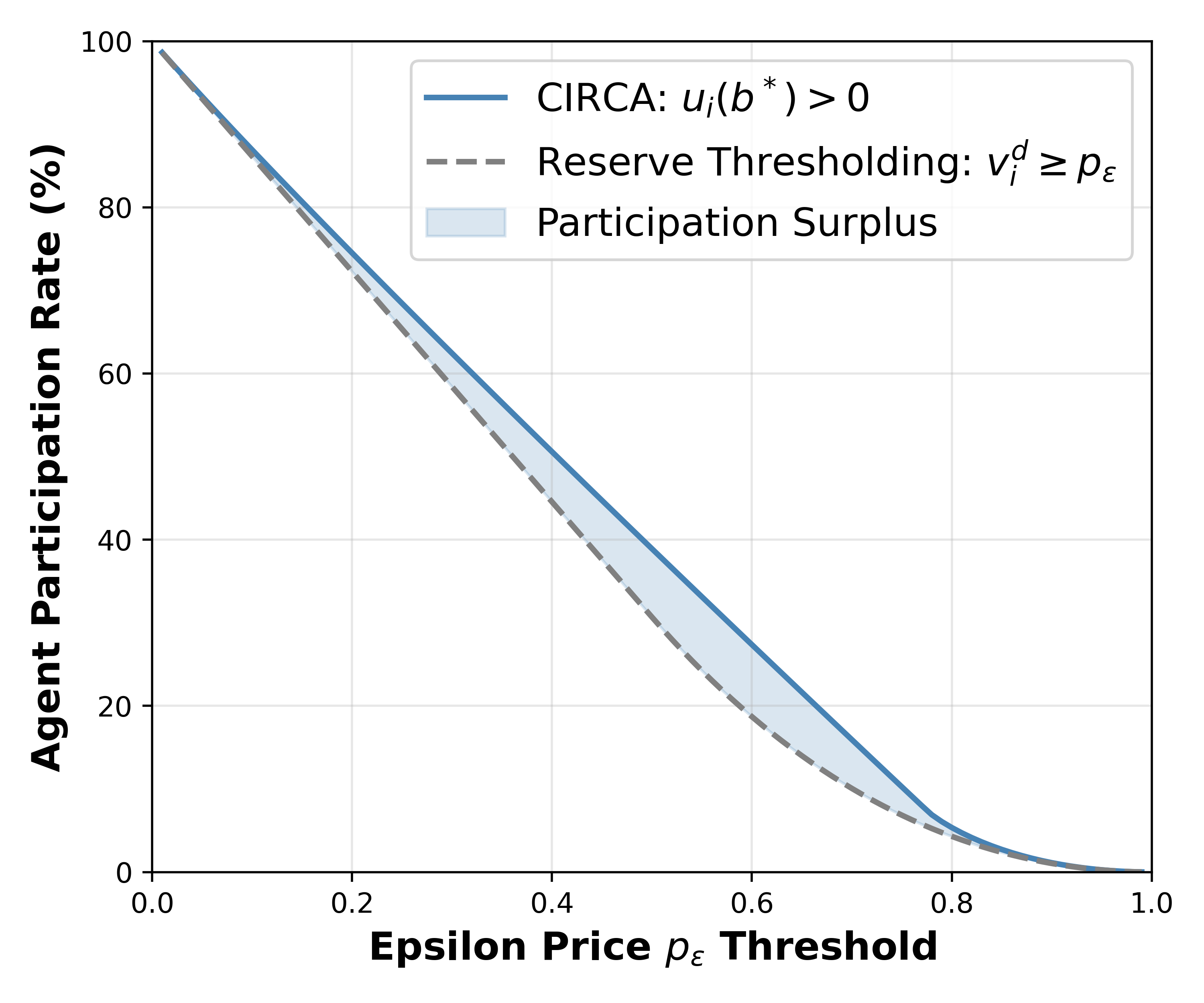}
    }
    \subfigure{
    \includegraphics[width=0.44\textwidth]{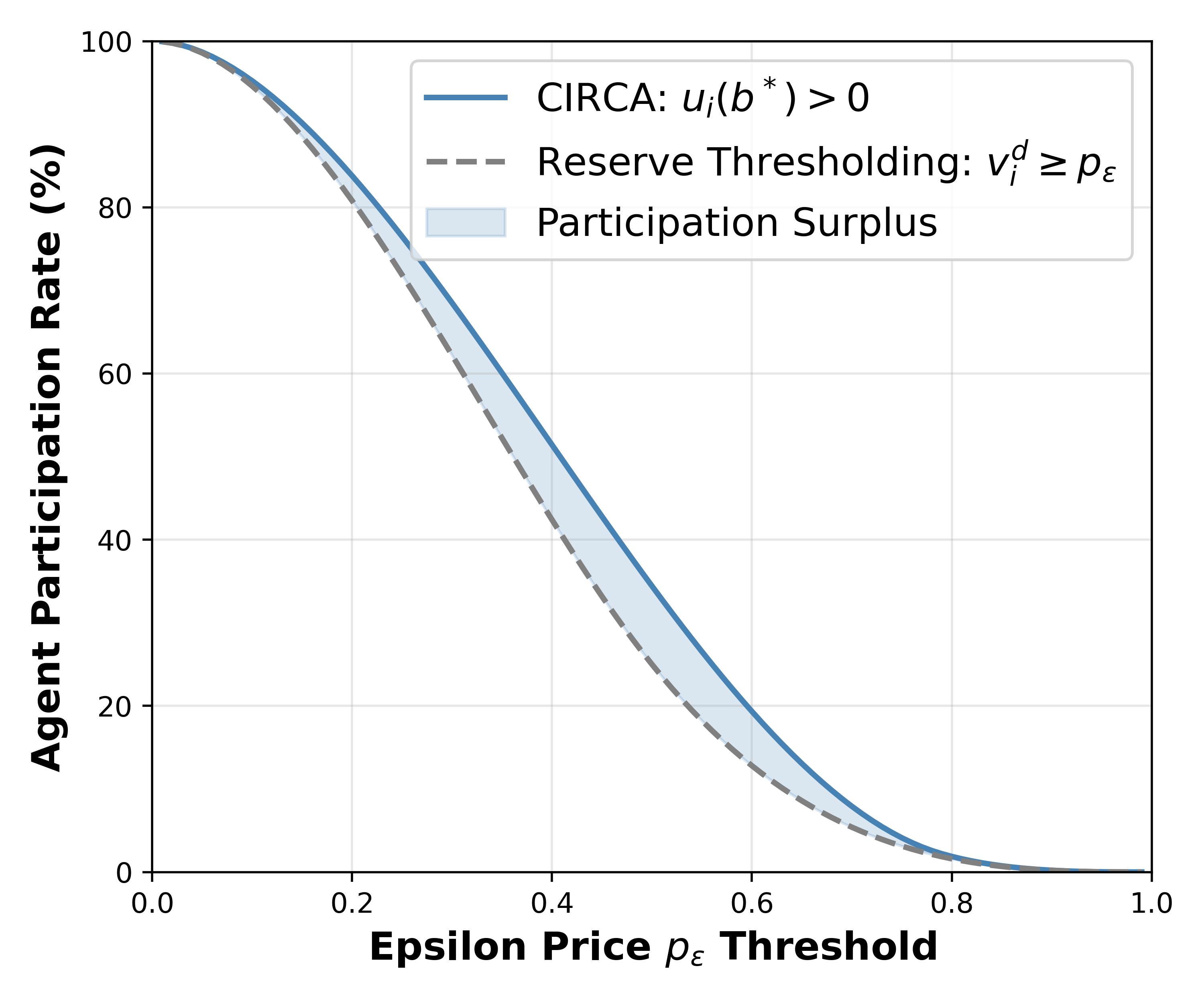}
    }
    \caption{
    \textbf{Improved Participation with Uniform \& Beta Values}. 
    When total value stems from a (left) Uniform $V_i \sim U(0,1)$ or (right) Beta distribution $V_i \sim \text{Beta}(\alpha=\beta=2)$, agents participate at a higher rate in \textsc{Circa} than Reserve Thresholding. \vspace{-4mm}
    }
    \label{fig:beta-uniform-participation}
\end{figure*}

\begin{figure*}[!tbp]
    \centering
    \subfigure{
    \includegraphics[width=0.44\textwidth]{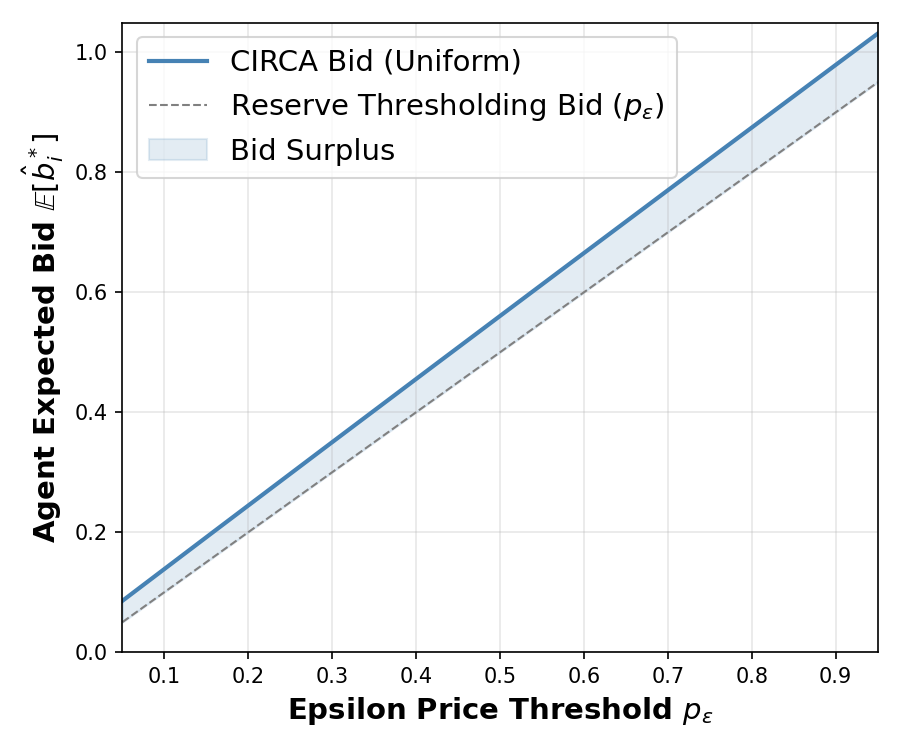}
    }
    \subfigure{
    \includegraphics[width=0.44\textwidth]{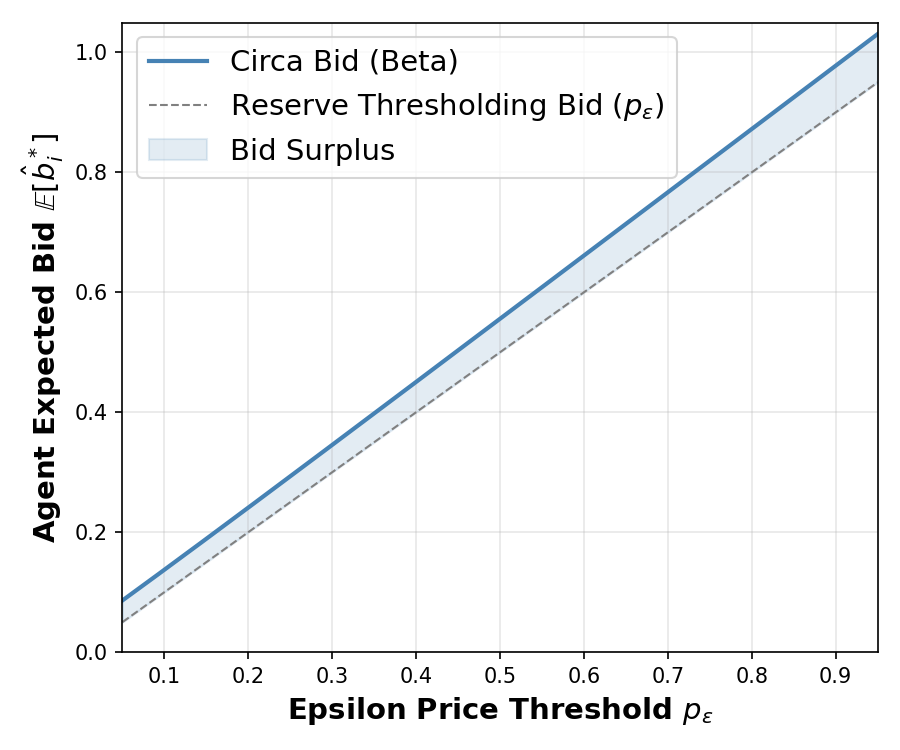}
    }
    \caption{
    \textbf{Improved Compliance with Uniform \& Beta Values}. 
    When total value stems from a (left) Uniform $V_i \sim U(0,1)$ or (right) Beta distribution $V_i \sim \text{Beta}(\alpha=\beta=2)$, agents bid more compliant models in \textsc{Circa} than Reserve Thresholding.
    }
    \label{fig:beta-uniform-bid}
\end{figure*}

Agents participate when they can gain positive utility. For our Reserve Thresholding baseline this is simple, an agent $i$ participates when their deployment value outstrips the price of compliance $v_i^d > p_\epsilon$. 
Thus, the participation rate for Reserve Thresholding is simply this probability $\mathbb{P}(v_i^d > p_\epsilon)$ across varying $p_\epsilon$.
In \textsc{Circa}, this probability, via Equation \ref{eq:bidding-strategy}, is $\mathbb{P}(v_i^d + \int_0^{v_i^p} F(z)dz > p_\epsilon)$ across varying $p_\epsilon$.
We approximate these probabilities for both Uniform and Beta(2,2) distributions and plot the results in Figure \ref{fig:beta-uniform-participation}.
As expected, one can see that \textsc{Circa} achieves a greater participation rate, by upwards of 15\%, compared to the Reserve Thresholding baseline across all values of $p_\epsilon$.
We note that participation rates are similar between the two methods for very low or high values of $p_\epsilon$. 
The reason is simple. 
As $p_\epsilon \rightarrow 0$, all agents will achieve positive utility, and thus participate, irrespective of the mechanism.
Likewise, as $p_\epsilon \rightarrow 1$, no agents will be able to achieve positive utility, and thus no agents will participate irrespective of the mechanism.

To analyze the expected participating bid in \textsc{Circa}, we leverage the results of Proposition \ref{prop:expected-bid}.
Namely, we approximate the integral in Equation \ref{eq:expected-bid}, for both Uniform and Beta(2,2) distributions, to compute the expected agent bid for various values of $p_\epsilon$.
The results, provided in Figure \ref{fig:beta-uniform-bid}, show that \textsc{Circa} incentivizes agents to bid more than they would under our Reserve Thresholding baseline, by upwards of 20\%, across all values of $p_\epsilon$.

\begin{figure}[!htbp]
    \centering
    \begin{minipage}{0.45\textwidth}
        \centering
        \captionof{table}{Equalized Odds as Minority Class Data Increases.}
        \label{tab:ablation}
        \begin{tabular}{cc}
            \toprule
            \textbf{Minority Class \%} & \textbf{Mean EO Score} \\ 
            \midrule
            5\%  & 22.55 \\
            10\% & 22.31 \\
            15\% & 18.97 \\
            20\% & 17.46 \\
            25\% & 15.78 \\
            30\% & 15.44 \\
            35\% & 13.09 \\
            40\% & 11.01 \\
            45\% & 9.83  \\
            50\% & 9.38  \\
            \bottomrule
        \end{tabular}
    \end{minipage}
    \hfill
    \begin{minipage}{0.5\textwidth}
        \centering
        \includegraphics[width=\linewidth]{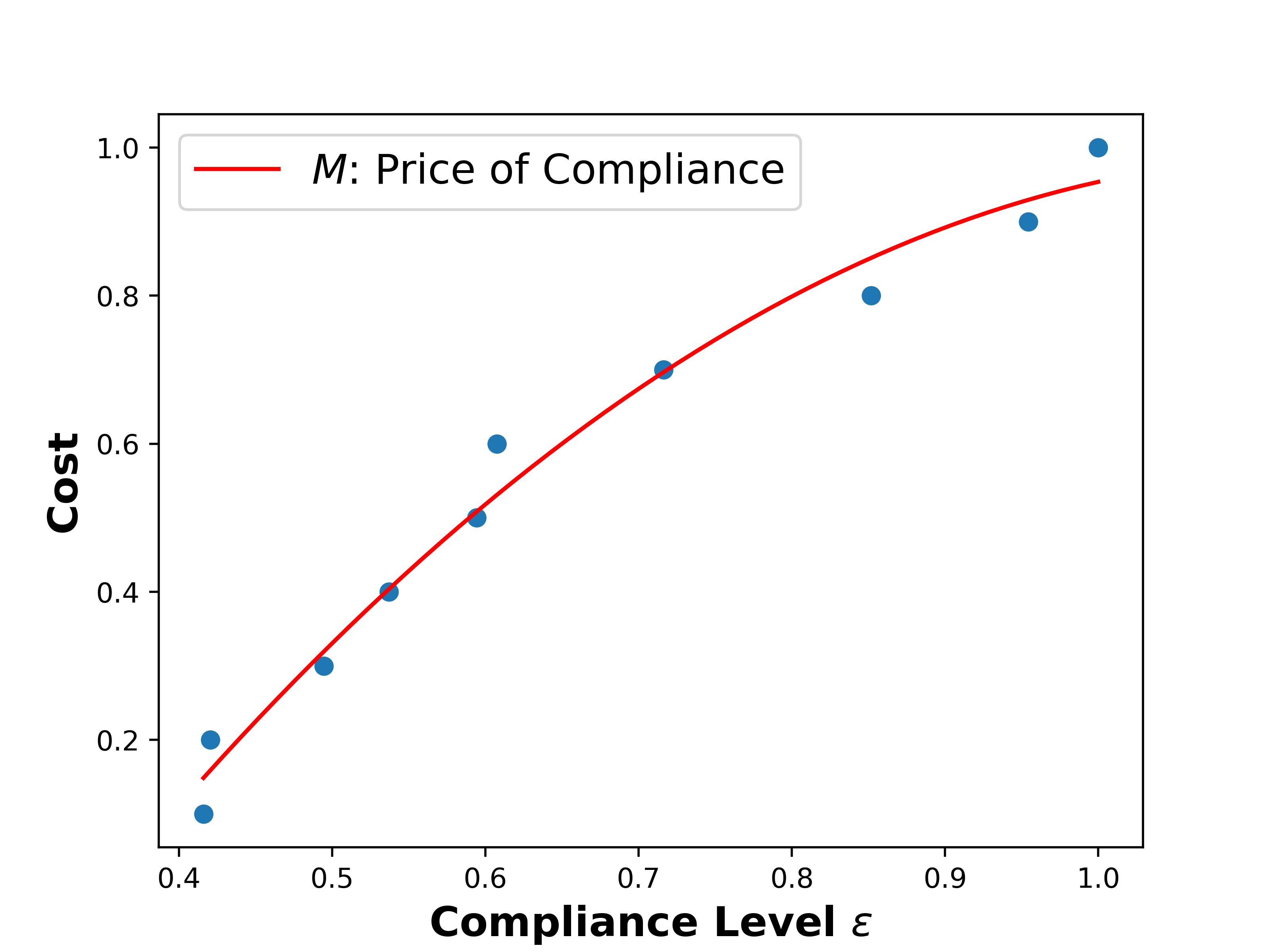}
        \caption{\textbf{Strictly Monotonic Compliance-Cost Relationship.} As the percentage of minority class data increases (greater cost), equalized odds metric improves (greater compliance) on Fairface.}
        \label{fig:ablation}
    \end{minipage}
\end{figure}

\textbf{Compliance-Cost Case Study}.
Finally, in our third experiment, a case study is conducted to demonstrate that in realistic settings, compliance is mapped to cost in a monotonically increasing way (as detailed in Assumption \ref{assumption:a2}).
While there are many compliance metrics to consider when gauging AI deployment, model fairness is analyzed, via equalized odds, for image classification in this study. 
Equalized odds measures if different groups have similar true positive rates and false positive rates (lower is better).
Multiple VGG-16 models are trained on the Fairface dataset \citep{karkkainen2021fairface} for fifty epochs (repeated ten times with different random seeds), and consider a gender classification task with race as the sensitive attribute.
Models with the largest validation classification accuracy during training are selected for testing.
Many types of costs exist for training compliant models, such as extensive architecture and hyper-parameter search. 
In this study, the cost of an agent acquiring more minority class data is considered. 
Acquiring more minority class data leads to a larger and more balanced dataset. 
Various mixtures of training data are simulated, starting from a 95:5 skew and scaling up to fully balanced training data with respect to the sensitive attribute. 
In this study, equalized odds performance is gauged on well-balanced test data for the models trained on various mixtures of data. 
The results of this case study are shown in Figure \ref{fig:ablation} and Table \ref{tab:ablation}.

As expected, in Table \ref{tab:ablation}, the equalized odds score decreases (more compliant model) when collecting more minority class data (increased cost).
To adjust equalized odds to fit into the setting where $\epsilon \in (0,1)$, the original equalized odds score is inverted and normalized.
In Figure \ref{fig:ablation}, one can see that compliance level is indeed monotonically increasing with respect to the cost.

\section{Conclusion}
\label{sec:conclusion}

As AI models grow, the risks associated with their misuse become significant, particularly given their opaque, black-box nature. 
Establishing robust algorithmic safeguards is crucial to protect users from unethical, unsafe, or illegally-deployed models. 
In this paper, we present a regulatory framework designed to ensure that only models deemed compliant by a regulator can be deployed for public use.
Our key contribution is the development of an auction-based regulatory mechanism that simultaneously (i) enforces compliance standards and (ii) provably incentivizes agents to exceed minimum compliance thresholds. 
This approach encourages broader participation and the development of more compliant models compared to baseline regulatory methods. 
Empirical results confirm that our mechanism increases agent participation by 15\% and raises agent spending on compliance by 20\%, demonstrating its effectiveness at promoting more compliant AI deployment.

\section*{Ethical Considerations Statement}
Unchecked AI deployment runs the risk of unsafe consequences that can harm users and stoke division within our society.
It is imperative to outline and employ regulatory frameworks to mitigate these dangers and ensure user safety.
However, regulation in AI is heavily under-researched.
The goal of this paper is to take a step towards designing realistic and effective regulation to ensure AI model compliance.
We hope that the impact of our paper will spur future research into regulatory AI, and soon provide a robust solution for governments to implement.

\section*{Acknowledgements}
Bornstein, Che, and Huang are supported by DARPA Transfer from Imprecise and Abstract Models to Autonomous
Technologies (TIAMAT) 80321, National Science Foundation NSF-IIS-2147276 FAI, DOD-ONR-Office of Naval
Research under award number N00014-22-1-2335, DOD-AFOSR-Air Force Office of Scientific Research under award
number FA9550-23-1-0048, DOD-DARPA-Defense Advanced Research Projects Agency Guaranteeing AI Robustness
against Deception (GARD) HR00112020007, Adobe, Capital One and JP Morgan faculty fellowships.

\bibliography{bibliography}

@misc{bhaskar18,
  author        = {V. Bhaskar},
  title         = {Lecture 8: All Pay Auction},
  month         = {January},
  year          = {2018},
  publisher={University College London}
}

@misc{tardos17,
  author        = {Eva Tardos},
  title         = {Lecture 17: All-pay Auctions},
  month         = {March},
  year          = {2017},
  publisher={Cornell University},
  url={https://www.cs.cornell.edu/courses/cs6840/2017sp/lecnotes/lec17.pdf}
}

@article{amann1996asymmetric,
  title={Asymmetric all-pay auctions with incomplete information: the two-player case},
  author={Amann, Erwin and Leininger, Wolfgang},
  journal={Games and economic behavior},
  volume={14},
  number={1},
  pages={1--18},
  year={1996},
  publisher={Elsevier}
}

@article{barman2024,
title = {The Dark Side of Language Models: Exploring the Potential of LLMs in Multimedia Disinformation Generation and Dissemination},
journal = {Machine Learning with Applications},
volume = {16},
pages = {100545},
year = {2024},
issn = {2666-8270},
doi = {https://doi.org/10.1016/j.mlwa.2024.100545},
url = {https://www.sciencedirect.com/science/article/pii/S2666827024000215},
author = {Dipto Barman and Ziyi Guo and Owen Conlan},
}

@misc{sun2024exploring,
  title={Exploring the Deceptive Power of LLM-Generated Fake News: A Study of Real-World Detection Challenges},
  author={Sun, Yanshen and He, Jianfeng and Cui, Limeng and Lei, Shuo and Lu, Chang-Tien},
  journal={arXiv preprint arXiv:2403.18249},
  year={2024}
}

@misc{neumann2024diverse,
  title={Diverse, but Divisive: LLMs Can Exaggerate Gender Differences in Opinion Related to Harms of Misinformation},
  author={Neumann, Terrence and Lee, Sooyong and De-Arteaga, Maria and Fazelpour, Sina and Lease, Matthew},
  journal={arXiv preprint arXiv:2401.16558},
  year={2024}
}

@misc{jacob2024, 
author  = {Jacob, Denny},
title={OpenAI forms New Committee to Evaluate Safety, Security}, 
url={https://www.wsj.com/tech/ai/openai-forms-new-committee-to-evaluate-safety-security-4a6e74bb}, 
journal={The Wall Street Journal},
date={2024-05-28},
year={2024}}

@misc{bruell2023, 
author  = {Bruell, Alexandra},
title={New York Times Sues Microsoft and OpenAI, Alleging Copyright Infringement}, 
url={https://www.wsj.com/tech/ai/new-york-times-sues-microsoft-and-openai-alleging-copyright-infringement-fd85e1c4}, 
journal={The Wall Street Journal}, 
date={2023-12-27},
year={2023}}

@misc{robertson2024,
 author  = {Robertson, Adi},
 title   = {Google apologizes for ‘missing the mark’ after Gemini generated racially diverse Nazis},
 journal = {The Verge},
 url     = {https://www.theverge.com/2024/2/21/24079371/google-ai-gemini-generative-inaccurate-historical},
 date = {2024-02-21},
year={2024}
}

@misc{moreno2023,
 author  = {Moreno, J. Edward},
 title   = {Boom in A.I. Prompts a Test of Copyright Law},
 journal = {The New York Times},
 url  =  {https://www.nytimes.com/2023/12/30/business/media/copyright-law-ai-media.html},
 date = {2023-12-30},
year={2023}
}

@misc{white2023,
 author  = {White, Jeremy},
 title   = {How Strangers Got My Email Address From ChatGPT’s Model},
 journal = {The New York Times},
 url     = {https://www.nytimes.com/interactive/2023/12/22/technology/openai-chatgpt-privacy-exploit.html},
 date = {2023-12-22},
year={2023}
}

@misc{white2024,
 author  = {White, Jeremy},
 title   = {See How Easily A.I. Chatbots Can Be Taught to Spew Disinformation},
 journal = {The New York Times},
 url     = {https://www.nytimes.com/interactive/2024/05/19/technology/biased-ai-chatbots.html},
 date = {2024-05-19},
year={2024}
}

@misc{metz2024,
 author  = {Metz, Cade and Kang, Cecilia and Frenkel, Sheera and Thompson, Stuart A. and Grant, Nico},
 title   = {How Tech Giants Cut Corners to Harvest Data for A.I.},
 journal = {The New York Times},
 url     = {https://www.nytimes.com/2024/04/06/technology/tech-giants-harvest-data-artificial-intelligence.html},
 date = {2024-04-06},
year={2024}
}

@misc{seetharaman2024, 
author  = {Seetharaman, Deepa},
title={OpenAI, Meta and Google Sign On to New Child Exploitation Safety Measures}, 
url={https://www.wsj.com/tech/ai/ai-developers-agree-to-new-safety-measures-to-fight-child-exploitation-2a58129c}, 
journal={The Wall Street Journal}, 
date={2024-04-23},
year={2024}}

@misc{brewster2024, 
author  = {Brewster, Jack},
title={How I Built an AI-Powered, Self-Running Propaganda Machine for \$105}, 
url={https://www.wsj.com/politics/how-i-built-an-ai-powered-self-running-propaganda-machine-for-105-e9888705}, 
journal={The Wall Street Journal}, 
date={2024-04-12},
year={2024}}

@misc{whitehouse2023,
	author = {The White House},
	title = {FACT SHEET: President Biden Issues Executive Order on Safe, Secure, and Trustworthy Artificial Intelligence},
	year = {2023},
}

@misc{cabill2024,
	author = {California Legislative Information},
	title = {SB-1047 Safe and Secure Innovation for Frontier Artificial Intelligence Models Act.},
	howpublished = {\url{https://leginfo.legislature.ca.gov/faces/billNavClient.xhtml?bill_id=202320240SB1047}},
	year = {2024},
}

@misc{yaghini2024regulation,
  title={Regulation Games for Trustworthy Machine Learning},
  author={Yaghini, Mohammad and Liu, Patty and Boenisch, Franziska and Papernot, Nicolas},
  journal={arXiv preprint arXiv:2402.03540},
  year={2024}
}

@article{rodriguez2022collusion,
  title={Collusion detection in public procurement auctions with machine learning algorithms},
  author={Rodr{\'\i}guez, Manuel J Garc{\'\i}a and Rodr{\'\i}guez-Montequ{\'\i}n, Vicente and Ballesteros-P{\'e}rez, Pablo and Love, Peter ED and Signor, Regis},
  journal={Automation in Construction},
  volume={133},
  pages={104047},
  year={2022},
  publisher={Elsevier}
}

@article{de2021artificial,
  title={Artificial intelligence regulation: a framework for governance},
  author={de Almeida, Patricia Gomes R{\^e}go and dos Santos, Carlos Denner and Farias, Josivania Silva},
  journal={Ethics and Information Technology},
  volume={23},
  number={3},
  pages={505--525},
  year={2021},
  publisher={Springer}
}

@misc{jagadeesan2024safety,
  title={Safety vs. Performance: How Multi-Objective Learning Reduces Barriers to Market Entry},
  author={Jagadeesan, Meena and Jordan, Michael I and Steinhardt, Jacob},
  journal={arXiv preprint arXiv:2409.03734},
  year={2024}
}

@article{baye1996all,
  title={The all-pay auction with complete information},
  author={Baye, Michael R and Kovenock, Dan and De Vries, Casper G},
  journal={Economic Theory},
  volume={8},
  pages={291--305},
  year={1996},
  publisher={Springer}
}

@misc{dipalantino2009crowdsourcing,
  title={Crowdsourcing and all-pay auctions},
  author={DiPalantino, Dominic and Vojnovic, Milan},
  booktitle={Proceedings of the 10th ACM conference on Electronic commerce},
  pages={119--128},
  year={2009}
}

@article{siegel2009all,
  title={All-pay contests},
  author={Siegel, Ron},
  journal={Econometrica},
  volume={77},
  number={1},
  pages={71--92},
  year={2009},
  publisher={Wiley Online Library}
}

@misc{goeree2000all,
  title={All-pay-all auctions},
  author={Goeree, Jacob K and Turner, John L},
  journal={University of Virginia, Mimeo},
  year={2000}
}

@article{gemp2022designing,
  title={Designing all-pay auctions using deep learning and multi-agent simulation},
  author={Gemp, Ian and Anthony, Thomas and Kramar, Janos and Eccles, Tom and Tacchetti, Andrea and Bachrach, Yoram},
  journal={Scientific Reports},
  volume={12},
  number={1},
  pages={16937},
  year={2022},
  publisher={Nature Publishing Group UK London}
}

@misc{karkkainen2021fairface,
  title={Fairface: Face attribute dataset for balanced race, gender, and age for bias measurement and mitigation},
  author={Karkkainen, Kimmo and Joo, Jungseock},
  booktitle={Proceedings of the IEEE/CVF winter conference on applications of computer vision},
  pages={1548--1558},
  year={2021}
}

@InProceedings{N18-1101,
  author = "Williams, Adina
            and Nangia, Nikita
            and Bowman, Samuel",
  title = "A Broad-Coverage Challenge Corpus for Sentence Understanding through Inference",
  booktitle = "Proceedings of the 2018 Conference of the North American Chapter of the Association for Computational Linguistics: Human Language Technologies, Volume 1 (Long Papers)",
  year = "2018",
  publisher = "Association for Computational Linguistics",
  pages = "1112--1122",
  address = "New Orleans, Louisiana",
  url = "http://aclweb.org/anthology/N18-1101"
}

@inproceedings{rajpurkar-etal-2016-squad,
    title = "{SQ}u{AD}: 100,000+ Questions for Machine Comprehension of Text",
    author = "Rajpurkar, Pranav  and
      Zhang, Jian  and
      Lopyrev, Konstantin  and
      Liang, Percy",
    editor = "Su, Jian  and
      Duh, Kevin  and
      Carreras, Xavier",
    booktitle = "Proceedings of the 2016 Conference on Empirical Methods in Natural Language Processing",
    month = nov,
    year = "2016",
    address = "Austin, Texas",
    publisher = "Association for Computational Linguistics",
    url = "https://aclanthology.org/D16-1264/",
    doi = "10.18653/v1/D16-1264",
    pages = "2383--2392"
}

@misc{FDA_E19_2019,
  author = {{U.S. Food and Drug Administration}},
  title = {E19 A Selective Approach to Safety Data Collection in Specific Late-Stage Pre-Approval or Post-Approval Clinical Trials},
  year = {2022},
  url = {https://www.fda.gov/regulatory-information/search-fda-guidance-documents/e19-selective-approach-safety-data-collection-specific-late-stage-pre-approval-or-post-approval}
}

@misc{wang2023decodingtrust,
  title={DecodingTrust: A Comprehensive Assessment of Trustworthiness in GPT Models.},
  author={Wang, Boxin and Chen, Weixin and Pei, Hengzhi and Xie, Chulin and Kang, Mintong and Zhang, Chenhui and Xu, Chejian and Xiong, Zidi and Dutta, Ritik and Schaeffer, Rylan and others},
  booktitle={NeurIPS},
  year={2023}
}

@misc{munoz2024pyrit,
  title={PyRIT: A Framework for Security Risk Identification and Red Teaming in Generative AI System},
  author={Munoz, Gary D Lopez and Minnich, Amanda J and Lutz, Roman and Lundeen, Richard and Dheekonda, Raja Sekhar Rao and Chikanov, Nina and Jagdagdorj, Bolor-Erdene and Pouliot, Martin and Chawla, Shiven and Maxwell, Whitney and others},
  journal={arXiv preprint arXiv:2410.02828},
  year={2024}
}

@misc{howe2024effects,
  title={Effects of Scale on Language Model Robustness},
  author={Howe, Nikolaus and McKenzie, Ian and Hollinsworth, Oskar and Zajac, Micha{\l} and Tseng, Tom and Tucker, Aaron and Bacon, Pierre-Luc and Gleave, Adam},
  journal={arXiv preprint arXiv:2407.18213},
  year={2024}
}

@misc{liu2024towards,
  title={Towards safer large language models through machine unlearning},
  author={Liu, Zheyuan and Dou, Guangyao and Tan, Zhaoxuan and Tian, Yijun and Jiang, Meng},
  journal={arXiv preprint arXiv:2402.10058},
  year={2024}
}

@misc{daisafe,
  title={Safe RLHF: Safe Reinforcement Learning from Human Feedback},
  author={Dai, Josef and Pan, Xuehai and Sun, Ruiyang and Ji, Jiaming and Xu, Xinbo and Liu, Mickel and Wang, Yizhou and Yang, Yaodong},
  booktitle={The Twelfth International Conference on Learning Representations},
  year={2024}
}

@article{stavins2008meaningful,
  title={A meaningful US cap-and-trade system to address climate change},
  author={Stavins, Robert N},
  journal={Harv. Envtl. L. Rev.},
  volume={32},
  pages={293},
  year={2008},
  publisher={HeinOnline}
}

@article{goulder2013carbon,
  title={Carbon taxes versus cap and trade: a critical review},
  author={Goulder, Lawrence H and Schein, Andrew R},
  journal={Climate Change Economics},
  volume={4},
  number={03},
  pages={1350010},
  year={2013},
  publisher={World Scientific}
}

@article{ouyang2022training,
  title={Training language models to follow instructions with human feedback},
  author={Ouyang, Long and Wu, Jeffrey and Jiang, Xu and Almeida, Diogo and Wainwright, Carroll and Mishkin, Pamela and Zhang, Chong and Agarwal, Sandhini and Slama, Katarina and Ray, Alex and others},
  journal={Advances in neural information processing systems},
  volume={35},
  pages={27730--27744},
  year={2022}
}

@misc{christiano2017deep,
  title={Deep reinforcement learning from human preferences},
  author={Christiano, Paul F and Leike, Jan and Brown, Tom and Martic, Miljan and Legg, Shane and Amodei, Dario},
  journal={Advances in neural information processing systems},
  volume={30},
  year={2017}
}

@misc{kaufmann2023survey,
  title={A survey of reinforcement learning from human feedback},
  author={Kaufmann, Timo and Weng, Paul and Bengs, Viktor and H{\"u}llermeier, Eyke},
  journal={arXiv preprint arXiv:2312.14925},
  year={2023}
}

@misc{act2024eu,
  title={The eu artificial intelligence act},
  author={Act, EU Artificial Intelligence},
  journal={European Union},
  year={2024}
}

@misc{trading_inference_time_compute,
  title={Trading Inference-Time Compute for Adversarial Robustness},
  author={Zaremba, Wojciech and Nitishinskaya, Evgenia and Barak, Boaz and Lin, Stephanie and Toyer, Sam and Yu, Yaodong and Dias, Rachel and Wallace, Eric and Xiao, Kai and Heidecke, Johannes and Glaese, Amelia},
  year={2025},
  howpublished={\url{https://cdn.openai.com/papers/trading-inference-time-compute-for-adversarial-robustness-20250121_1.pdf}},
}

@article{huang2025survey,
  title={A survey on hallucination in large language models: Principles, taxonomy, challenges, and open questions},
  author={Huang, Lei and Yu, Weijiang and Ma, Weitao and Zhong, Weihong and Feng, Zhangyin and Wang, Haotian and Chen, Qianglong and Peng, Weihua and Feng, Xiaocheng and Qin, Bing and others},
  journal={ACM Transactions on Information Systems},
  volume={43},
  number={2},
  pages={1--55},
  year={2025},
  publisher={ACM New York, NY}
}

@book{powell2014science,
  title={Science at EPA: Information in the regulatory process},
  author={Powell, Mark R},
  year={2014},
  publisher={Routledge},
  address={New York, NY}
}

@book{coglianese2007regulation,
  title={Regulation and Regulatory Processes},
  author={Coglianese, C. and Kagan, R.A.},
  isbn={9780754625186},
  lccn={2007923516},
  series={International library of essays in law and society},
  url={https://books.google.com/books?id=C0RnNQAACAAJ},
  year={2007},
  publisher={Ashgate},
  address={Aldershot, England}
}

@book{krishna2009auction,
  title={Auction theory},
  author={Krishna, Vijay},
  year={2009},
  publisher={Academic Press},
  address={Burlington, MA}
}

@techreport{vanschoren2025role,
  title       = {The Role of {AI} Safety Benchmarks in Evaluating Systemic Risks
                 in General-Purpose {AI} Models},
  author      = {Vanschoren, Joaquin and
                 Fern{\'a}ndez Llorca, David and
                 Eriksson, Maria and
                 G{\'o}mez, Emilia},
  institution = {European Commission, Joint Research Centre},
  year        = {2025},
  month       = {October},
  number      = {JRC143259},
  doi         = {10.2760/1807342},
  url         = {https://publications.jrc.ec.europa.eu/repository/handle/JRC143259},
  address     = {Luxembourg},
  publisher   = {Publications Office of the European Union},
  isbn        = {978-92-68-31428-9}
}

@techreport{fli2025aisafetyindex,
  title        = {AI Safety Index: Winter 2025},
  author       = {{Future of Life Institute}},
  institution  = {Future of Life Institute},
  year         = {2025},
  month        = {December},
  url          = {https://futureoflife.org/wp-content/uploads/2025/12/AI-Safety-Index-Report_131225_Full_Report_Digital.pdf},
  note         = {Third edition. Independent assessment of eight leading AI companies' safety practices.}
}

@techreport{aisi2025frontier,
  title       = {Frontier {AI} Trends Report},
  author      = {{AI Security Institute}},
  institution = {Department for Science, Innovation and Technology,
                 UK Government},
  year        = {2025},
  month       = {December},
  url         = {https://www.aisi.gov.uk/frontier-ai-trends-report},
  note        = {First public analysis drawing on two years of evaluations
                 of over 30 frontier AI models across cyber, biology,
                 chemistry, and autonomous capabilities.}
}

@misc{bengio2025international,
  title   = {International {AI} Safety Report},
  author  = {Bengio, Yoshua and Mindermann, S{\"o}ren and Privitera, Daniel
             and Besiroglu, Tamay and Bommasani, Rishi and Casper, Stephen
             and Choi, Yejin and Fox, Philip and Garfinkel, Ben and
             Goldfarb, Danielle and Heidari, Hoda and Ho, Anson and
             Kapoor, Sayash and Khalatbari, Leila and Longpre, Shayne and
             Manning, Sam and Mavroudis, Vasilios and Mazeika, Mantas and
             Michael, Julian and Newman, Jessica and others},
  journal = {arXiv preprint arXiv:2501.17805},
  year    = {2025},
  month   = {January},
  doi     = {10.48550/arXiv.2501.17805},
  url     = {https://arxiv.org/abs/2501.17805},
  note    = {DSIT 2025/001. Mandated by 30 nations, the UN, OECD, and EU
             at the Bletchley AI Safety Summit. Led by Yoshua Bengio with
             96 contributing experts.}
}

@misc{laufer2025backfiring,
  title={The Backfiring Effect of Weak AI Safety Regulation},
  author={Laufer, Benjamin and Kleinberg, Jon and Heidari, Hoda},
  journal={arXiv preprint arXiv:2503.20848},
  year={2025}
}

@inproceedings{yew2024,
author = {Yew, Rui-Jie},
title = {Break It 'Til You Make It: An Exploration of the Ramifications of Copyright Liability Under a Pre-training Paradigm of AI Development},
year = {2024},
isbn = {9798400703331},
publisher = {Association for Computing Machinery},
address = {New York, NY, USA},
url = {https://doi.org/10.1145/3614407.3643707},
doi = {10.1145/3614407.3643707},
booktitle = {Proceedings of the 2024 Symposium on Computer Science and Law},
pages = {64–72},
numpages = {9},
location = {Boston, MA, USA},
series = {CSLAW '24}
}

@misc{qiu2025modeling,
  title={Modeling the Economic Impacts of AI Openness Regulation},
  author={Qiu, Tori and Laufer, Benjamin and Kleinberg, Jon and Heidari, Hoda},
  journal={arXiv preprint arXiv:2507.14193},
  year={2025}
}

@misc{xu2025economics,
  title={The economics of ai foundation models: Openness, competition, and governance},
  author={Xu, Fasheng and Wang, Xiaoyu and Chen, Wei and Xie, Karen},
  journal={arXiv preprint arXiv:2510.15200},
  year={2025}
}

@article{bertoletti2016reserve,
  title={Reserve prices in all-pay auctions with complete information},
  author={Bertoletti, Paolo},
  journal={Research in Economics},
  volume={70},
  number={3},
  pages={446--453},
  year={2016},
  publisher={Elsevier}
}

@article{kaplan2002all,
  title={All--pay auctions with variable rewards},
  author={Kaplan, Todd and Luski, Israel and Sela, Aner and Wettstein, David},
  journal={The Journal of Industrial Economics},
  volume={50},
  number={4},
  pages={417--430},
  year={2002},
  publisher={Wiley Online Library}
}

@misc{cfr42part84,
  title     = {Title 42 -- Public Health, Part 84 -- Approval of Respiratory Protective Devices},
  author    = {{Code of Federal Regulations}},
  year      = {1995},
  url       = {https://www.ecfr.gov/current/title-42/chapter-I/subchapter-G/part-84},
  note      = {Effective July 1995}
}

@misc{epa2024multipollutant,
  title  = {Multi-Pollutant Emissions Standards for Model Years 2027 and Later Light-Duty and Medium-Duty Vehicles},
  author = {{U.S. Environmental Protection Agency}},
  year   = {2024},
  note   = {Final Rule, 89 FR 27842}
}

@article{van2016drugs,
  title={Drugs, devices, and the FDA: part 2: an overview of approval processes: FDA approval of medical devices},
  author={Van Norman, Gail A},
  journal={JACC: Basic to Translational Science},
  volume={1},
  number={4},
  pages={277--287},
  year={2016},
  publisher={American College of Cardiology Foundation Washington, DC}
}

@misc{epanpdes,
  title  = {National Pollutant Discharge Elimination System ({NPDES})},
  author = {{U.S. Environmental Protection Agency}},
  year   = {2024},
  url    = {https://www.epa.gov/npdes}
}

@article{moldovanu2006contest,
  title={Contest architecture},
  author={Moldovanu, Benny and Sela, Aner},
  journal={Journal of Economic Theory},
  volume={126},
  number={1},
  pages={70--96},
  year={2006},
  publisher={Elsevier}
}
\bibliographystyle{plainnat}

\clearpage
{\begin{center} \bf \LARGE
    Appendix
    \end{center}
}
\appendix

\section{Notation Table}

\begin{table}[H]
\caption{Notating and Defining all Variables Listed Within \textsc{Circa}.}
\centering
\begin{tabular}{@{}>{\bfseries\centering\arraybackslash}p{0.8\linewidth} >{$}c<{$}@{}}
\toprule
\textbf{Definition} & \textbf{Notation} \\
\midrule
Regulator & R \\ \addlinespace
Number of Agents & n \\ \addlinespace
Compliance Threshold & \epsilon \\ \addlinespace
Compliance-to-Cost Function & M \\ \addlinespace
Price of Attaining Compliance & p_\epsilon \\ \addlinespace
Agent $i$ Bid & b_i \\ \addlinespace
Agent $i$'s Optimal Bid & b_i^* \\ \addlinespace
All Other Agents Bids & \bm{b}_{-i} \\ \addlinespace
Agent $i$ Utility & u_i \\ \addlinespace
Agent $i$ Model Compliance & s_i \\ \addlinespace
Total Value for Agent $i$ & V_i \\ \addlinespace
Total Value Distribution & \mathcal{D}_V \\ \addlinespace
Agent $i$ Scaling Factor & \lambda_i \\ \addlinespace
Scaling Factor Distribution & \mathcal{D}_\lambda \\ \addlinespace
Deployment Value for Agent $i$ & v_i^d \\ \addlinespace
Premium Compensation Value for Agent $i$ & v_i^p \\ \addlinespace
Probability Density Function for Premium Compensation & f_v \\ \addlinespace
Cumulative Distribution Function for Premium Compensation & F_v \\
\bottomrule
\end{tabular}
\end{table}

\section{Theoretical Proofs}
\label{app:proofs}

Below, we provide the full proofs of our Theorems and Corollaries presented within our work.

\subsection{Proof of Theorem 1}
\begin{reptheorem}{thm:simple-threshold}[Restated]
    Under Assumption \ref{assumption:a2}, agents participating in Reserve Thresholding Equation~\ref{eq:simple-threshold} have an optimal bid and utility of,
    $$
        b_i^* = p_\epsilon, \quad u_i(b_i^*; \; \bm{b}_{-i}, v_i^d) = v_i^d - p_\epsilon,
    $$
    and submit models with the following compliance level,
    $$
        s_i^* = \begin{cases}
            \epsilon &\text{ if } u_i(b_i^*; \; \bm{b}_{-i}, v_i^d) > 0,\\
            0 \text{ (no submission) } &\text{ else.}
        \end{cases}
    $$
\end{reptheorem}

\begin{proof}
From agent $i$'s utility within Reserve Thresholding, Equation~\ref{eq:simple-threshold}, it is clear that $u_i(0) = 0$.
We proceed to break the proof up into cases where agents have (1) a deployment value equal to or less than the price of compliance $p_\epsilon$ and (2) a deployment value larger than $p_\epsilon$.

\textbf{Case 1: $v_i^d \leq p_\epsilon$}. Leveraging  Equation~\ref{eq:simple-threshold}, when $b < p_\epsilon$ the indicator function returns $1_{(b \geq p_\epsilon)} = 0$. Therefore, agent utility is always negative when bidding less than $p_\epsilon$,
\begin{equation}
    u_i(b) = -b < 0.
\end{equation}
When $b \geq p_\epsilon$, the indicator function returns $1_{(b \geq p_\epsilon)} = 1$. Thus, agent utility becomes,
\begin{equation}
    u_i(b) = v_i^d - b < 0.
\end{equation}
Since this function is strictly decreasing with respect to $b$, the function is maximized at the smallest bid $b = p_\epsilon$,
\begin{equation}
    \label{eq:case1}
    b_i^* = \argmax_{b \in [p_\epsilon, 1]} u_i(b) = p_\epsilon \longrightarrow u_i(p_\epsilon) = v_i^d - p_\epsilon \leq p_\epsilon - p_\epsilon = 0.
\end{equation}
For an agent with deployment value at most equal to $p_\epsilon$, the upper bound on attainable utility when it participates, \textit{i.e.,} $b \in (0,1]$, is zero (Equation~\ref{eq:case1}).
Thus, agents have nothing to gain by participating, as they already start at zero utility $u_i(0) = 0$.
As a result, agents will not submit a model, $s_i^* = M(0) = 0$.

\textbf{Case 2: $v_i^d > p_\epsilon$}. By the same two-region argument as Case 1, bidding $b < p_\epsilon$ results in negative utility. Similarly, bidding $b \geq p_\epsilon$ results in an agent optimally bidding $b_i^* = p_\epsilon$. However, in Case 2, the agent utility becomes,
\begin{equation}
    \label{eq:case2}
    u_i(b_i^*) = v_i^d - p_\epsilon > 0.
\end{equation}
An agent with deployment value larger than $p_\epsilon$ will have a positive optimal utility when it participates (Equation~\ref{eq:case2}).
Furthermore, at this optimal bid, the corresponding compliance level is $s_i^* = M^{-1}(p_\epsilon) = \epsilon$.
\end{proof}

\subsection{Proof of Theorem 2}
\begin{reptheorem}{thm:general-bidding-strategy}[Restated]
    Agents participating in \prb{Circa} Equation~\ref{eq:all-pay-auction} follow an optimal bidding strategy $\hat{b}_i^*$ of,
    $$
        \hat{b}_i^* := p_\epsilon + v_i^p F(v_i^p) - \int_0^{v_i^p} F(z)dz \; > p_\epsilon, \quad u_i(\hat{b}_i^*; \; \bm{b}_{-i}, v_i^d, v_i^p) = v_i^d - p_\epsilon + \int_0^{v_i^p} F(z)dz,
    $$
    where $F(\cdot)$ denotes the cumulative distribution function (CDF) of the random premium reward variable corresponding to the premium reward $v_i^p = V_i\lambda_i$ with $F(v_i^p) > 0$ for $v_i^p>0$.
\end{reptheorem}

\begin{proof}
Before beginning our proof, we note that each agent $i$ cannot alter its own valuation $v_i^p$ for winning the all-pay auction.
Each valuation is private (unknown by other agents) and predetermined: total reward $V_i$ and partition factor $\lambda_i$ are randomly selected from a given distribution $\mathcal{D}$ on $[0,1]$ and $[0,1/2]$ respectively by ``nature''.
We define the cumulative distribution function for the auction reward $v_i^p = V_i \lambda_i$ as $F(\cdot)$ and the probability distribution function as $f(\cdot)$.
From Equation~\ref{eq:all-pay-auction}, we find that not participating (\textit{i.e.,} $b = 0$) results in no utility,
\begin{equation}
    \label{eq:zero-utility}
    u_i(0) = 0.
\end{equation}
An agent receives negative utility if its bid does not reach the price of compliance $p_\epsilon$,
\begin{equation}
    \label{eq:short-of-threshold}
    \max_{b \in (0, p_\epsilon)} u_i(b) < 0.
\end{equation}
Consequently, rational agents will either opt not to participate (notated as the set of agents $N$) or participate (notated as the set of agents $P$) and bid at least $p_\epsilon$.
We define these groups as,
\begin{align}
    N = \{ i \in [n] \; | \; \max_{b \in [0,1]} u_i(b) \leq 0 \}, \\
    \label{eq:participating-agents-1}
    P = \{ i \in [n] \; | \; \max_{b \in [0,1]} u_i(b) > 0 \}.
\end{align}
From here, we only focus on agents $i \in P$ which participate (\textit{i.e.,} have utility to be gained by participating).
As a result from Equations~\ref{eq:zero-utility} and \ref{eq:short-of-threshold}, Equation~\ref{eq:participating-agents-1} transforms into,
\begin{equation}
    \label{eq:participating-agents-2}
    P = \{ i \in [n] \; | \; \max_{b \in [p_\epsilon,1]} u_i(b) > 0 \}.
\end{equation}
The result of Equation~\ref{eq:participating-agents-2} is that participating agents bid at least $p_\epsilon$.
This is important, as every participating agent knows that all rival agents $j$ they will possibly be compared against have $b_j \in [p_\epsilon, 1]$.
Agents can dictate how much they bid, and we design our auction to ensure that agents bid in proportion to their valuation.
Following previous literature \citep{amann1996asymmetric, bhaskar18, tardos17, krishna2009auction}, we desire a \textit{monotone increasing equilibrium} bidding function $\beta(v^p): [0, 1/2] \rightarrow [p_\epsilon, 1]$. 
We conjecture that all rival agents $j \neq i$ follow this bidding function $\beta(v_j^p)$. We then derive agent $i$'s best response under this conjecture, and verify that the best response is $\beta(v_i^p)$, establishing Nash equilibrium by symmetry.
Using a bidding function transforms agent utility, for a given bid $b$, into the following,
\begin{align}
    u_i(b; \; v_i^d, v_i^p) &= \big( v_i^d + v_i^p \cdot 1_{(\text{if $i$ wins auction})} \big) \cdot \underbrace{1_{(\text{if } b \geq p_\epsilon)}}_{\text{satisfied for agents } i \in P} - \; \beta(b), \nonumber \\
     &= \mathbb{P}\big(\beta(b) > \beta(v_j^p) \big)v_i^p - \beta(b) + v_i^d, \quad b_j \sim \text{randomly sampled agent bid}.
\end{align}
Since $\beta(\cdot)$ is monotone increasing up to 1, and agents bidding $b = 1$ automatically win, one can see that $\mathbb{P}\big(\beta(b) > \beta(v_j^p) \big) = \mathbb{P}\big(b > v_j^p \big)$. Since $v_j^p \sim F$, it follows that $\mathbb{P}\big(b > v_j^p \big) = F(b)$. Thus, the utility function above becomes,
\begin{align}
    u_i(b; \; v_i^d, v_i^p) = v_i^p F(b) - \beta(b) + v_i^d.
\end{align}
At equilibrium, truthful reporting $b = v_i^p$ must be optimal. Differentiating with respect to $b$ and solving the first order conditions at $b = v_i^p$ yields,
\begin{equation}
    \frac{\partial u_i}{\partial b} \bigg|_{b = v_i^p} = v_i^p f(v_i^p) - \beta'(v_i^p) = 0.
\end{equation}
This yields the following ODE (as $v_i^p$ takes on values between 0 and 1/2),
\begin{equation}
    \beta'(v) = vf(v), \quad \text{for } v \in [0, 1/2].
\end{equation}
Now, solving the ODE at the specific value $v_i^p$,
\begin{equation}
    \int_0^{v_i^p}\beta'(v)dv = \int_{0}^{v_i^p}vf(v) dv.
\end{equation}
From Equation \ref{eq:participating-agents-2}, all participating agents bid at a minimum $\beta(0) = p_{\epsilon}$. Therefore, $\beta(0) = p_{\epsilon}$ is our boundary condition. Using integration by parts, and applying the boundary condition yields our desired result,
\begin{equation}
    \beta(v_i^p) - \beta(0) = vF(v)\big|_0^{v_i^p} - \int_0^{v_i^p}F(z)dz,
\end{equation}
\begin{equation}
    \hat{b}_i^* = \beta(v_i^p) := p_{\epsilon} + v_i^pF(v_i^p) - \int_0^{v_i^p}F(z)dz.
\end{equation}
Since $\int_0^{v_i^p}F(z)dz < \int_0^{v_i^p}F(v_i^p)dz = v_i^p F(v_i^p)$ for $v_i^p > 0$, it follows that,
\begin{equation}
    \hat{b}_i^* = p_{\epsilon} + v_i^pF(v_i^p) - \int_0^{v_i^p}F(z)dz > p_\epsilon.
\end{equation}
The general utility for \textsc{Circa} is detailed in Equation \ref{eq:regauc-utility-1} as $u_i(b; \; \bm{b}_{-i}, v_i^d, v_i^p) = v_i^d - p_\epsilon + v_i^p \cdot \bm{1}[\text{agent $i$ wins}]$. This is represented more formally as,
\begin{equation}
    u_i(b; \; \bm{b}_{-i}, v_i^d, v_i^p) = v_i^d - b + v_i^p F(v_i^p).
\end{equation}
Substituting in the optimal bid yields,
\begin{align}
    u_i(\hat{b}_i^*; \; \bm{b}_{-i}, v_i^d, v_i^p) &= v_i^d - \big(p_{\epsilon} + v_i^pF(v_i^p) - \int_0^{v_i^p}F(z)dz \big) + v_i^p F(v_i^p),\\
    &= v_i^d - p_{\epsilon} + \int_0^{v_i^p}F(z)dz.
\end{align}
\end{proof}

\subsection{Proof of Proposition 1}
\begin{repproposition}{prop:expected-bid}[Restated]
    Let $\hat{b}_i^*$ denote the optimal bidding strategy from Theorem~\ref{thm:general-bidding-strategy}. The expected participating agent's bid over the distribution of premium rewards $v_i^p$, where $F(\cdot)$ and $f(\cdot)$ denote the CDF and probability density function (PDF) of $v_i^p$, is,
    \begin{equation}
        \mathbb{E}[\hat{b}_i^*] = p_\epsilon + \int_0^{1/2} zf(z)(1 - F(z))dz \; > p_\epsilon.
    \end{equation}
\end{repproposition}

\begin{proof}
From Theorem~\ref{thm:general-bidding-strategy}, the optimal bid is,
\begin{equation}
    \hat{b}_i^* = p_\epsilon + v_i^p F(v_i^p) - \int_0^{v_i^p} F(z)\, dz.
\end{equation}
We simplify $v_i^p F(v_i^p) - \int_0^{v_i^p} F(z)\, dz$ using integration by parts. Recall that for any function $F$,
\begin{equation}
    \int_0^{v} z f(z)\, dz = \Big[z F(z)\Big]_0^{v} - \int_0^{v} F(z)\, dz = v F(v) - \int_0^{v} F(z)\, dz.
\end{equation}
Therefore,
\begin{equation}
    \hat{b}_i^* = p_\epsilon + \int_0^{v_i^p} z f(z)\, dz.
\end{equation}
Taking expectations over $v_i^p$, and applying the Law of the Unconscious Statistician, yields
\begin{equation}
    \mathbb{E}[\hat{b}_i^*] = p_\epsilon + \mathbb{E}\left[\int_0^{v_i^p} z f(z)\, dz\right] = p_\epsilon + \int_0^{1/2} \left(\int_0^{v} z f(z)\, dz\right) f(v)\, dv,
\end{equation}
where the outer $f(v)$ is the PDF of $v_i^p$ and the domain upper limit is $1/2$ since $v_i^p = V_i\lambda_i \leq 1/2$ by construction.
The double integral is taken over the region $\{(z, v) : 0 \leq z \leq v \leq 1/2\}$, which is a triangle in the $(z, v)$ plane with vertices $(0,0)$, $(0, 1/2)$, and $(1/2, 1/2)$. Since $zf(z)f(v) \geq 0$ on this region, we can switch the order of integration (via Fubini's theorem). For a fixed $z$, $v$ ranges over $[z, 1/2]$, giving,
\begin{equation}
    \mathbb{E}[\hat{b}_i^*] = p_\epsilon + \int_0^{1/2} z f(z) \left(\int_z^{1/2} f(v)\, dv\right) dz.
\end{equation}
The inner integral evaluates (since $F(1/2) = 1$ as $v_i^p \leq 1/2$) to,
\begin{equation}
    \int_z^{1/2} f(v)\, dv = F(1/2) - F(z) = 1 - F(z).
\end{equation}
This leads to our final result,
\begin{equation}
    \mathbb{E}[\hat{b}_i^*] = p_\epsilon + \int_0^{1/2} z f(z)(1 - F(z))\, dz > p_\epsilon.
\end{equation}
The final inequality holds since $z$, $f(z)$, and $(1- F(z))$ are positive for some $z \in (0, 1/2)$ (since $F(v_i^p) > 0$ for $v_i^p > 0$).

\end{proof}

\subsection{Proof of Corollary 1}
\begin{repcorollary}{cor:uniform-valuations}[Restated]
    Under Assumption \ref{assumption:a2}, for agents having total value $V_i$ and scaling factor $\lambda_i$ both stemming from a Uniform distribution, with $v_i^d = (1-\lambda_i)V_i,$ and $v_i^p = \lambda_iV_i$, their optimal bid and utility participating in \prb{Circa} (Equation~\ref{eq:all-pay-auction}) are,
    $$
    b_i^* := \min \{ \hat{b}_i^*, 1\}, \quad 
            \hat{b}_i^* = \begin{cases}
            p_\epsilon + \frac{(v_i^p)^2\ln(p_\epsilon)}{p_\epsilon-1} &\text{if } 0 \leq v_i^p \leq \frac{p_\epsilon}{2},\\
            p_\epsilon + \frac{8(v_i^p)^2(\ln(2v_i^p) - 1/2)+ p_\epsilon^2}{8(p_\epsilon-1)} &\text{if } \frac{p_\epsilon}{2} \leq v_i^p \leq \frac{1}{2},
        \end{cases}
    $$
    $$
        u_i(b_i^*; \; \bm{b}_{-i}, v_i^d, v_i^p) = 
        \begin{cases}
        \frac{2(v_i^p)^2\ln(p_\epsilon)}{p_\epsilon-1} + v_i^d -  b_i^* &\text{if } 0 \leq v_i^p \leq \frac{p_\epsilon}{2}, \\
        \frac{2(v_i^p)^2(\ln(2v_i^p)-1) + p_\epsilon}{p_\epsilon-1} + v_i^d -  b_i^* &\text{if } \frac{p_\epsilon}{2} \leq v_i^p \leq \frac{1}{2}.
        \end{cases}
    $$
    Participating agents submit models with compliance,
    $$
        s_i^* := \begin{cases}
             M^{-1}(b_i^*) > \epsilon &\text{ if } u_i(b_i^*; \; \bm{b}_{-i}, v_i^d, v_i^p) > 0,\\
            0 \text{ (no submission) } &\text{ else}.
        \end{cases}
    $$
\end{repcorollary}

\begin{proof}
    Let $v_i^p := V_i \lambda_i$, where $V_i \sim U[p_\epsilon, 1]$ and $\lambda_i \sim U[0,1/2]$.
    The reason that $V_i$ is within the interval $[p_\epsilon, 1]$, is that all participating agents must have a value of at least $p_\epsilon$ or else they would not have rationale to bid.
    The smallest value of $V_i$ such that this is possible is $p_\epsilon$, so it is the lower bound on this interval.
    Our first goal is to find the PDF of $v_i^p$, $f_{v_i^p}(\cdot)$. 

    We begin solving for $f_{v_i^p}(\cdot)$ by using a change of variables.
    For the product of two random variables $v = x_1 \cdot x_2$, let $y_1 = x_1 \cdot x_2$ and $y_2 = x_2$.
    Thus, we find inversely that $x_2 = y_2$ and $x_1 = y_1 / y_2$.
    Since $x_1$ and $x_2$ are independent and both uniform, we find that,
    \begin{equation}
        f_{y_1, y_2}(x_1, x_2) = (\frac{1}{1-p_\epsilon})(\frac{1}{1/2-0}) = \frac{2}{1-p_\epsilon}.
    \end{equation}
    When using the change of variables this becomes,
    \begin{equation}
        f_{y_1, y_2}(y_1, y_2) = f_{y_1, y_2}(x_1, x_2) |J|= \frac{2}{(1-p_\epsilon) y_2}, \quad |J| = \bigg| \begin{pmatrix}
            1/y_2 & -y_1/y_2^2 \\ 0 & 1
        \end{pmatrix}\bigg| = 1/y_2
    \end{equation}
    Marginalizing out $y_2$ (a non-negative value) yields,
    \begin{equation}
        f_{y_1}(y_1) = \int_{0}^{\infty} \frac{2}{(1-p_\epsilon) y_2} dy_2.
    \end{equation}
    The bounds of integration depend upon the value of $y_1$.
    The change of variable to the $(y_1, y_2)$ space, where $0 \leq y_1, y_2 \leq 1/2$, results in a new region of possible variable values.
    This region is a triangle bounded by the three vertices: $(0,0)$, $(p_\epsilon/2, 1/2)$, and $(1/2, 1/2)$.
    Thus, the bounds of marginalization depend upon the value of $y_1$. 
    For $0 \leq y_1 \leq p_\epsilon/2$ we have,
    \begin{equation}
        f_{y_1}(y_1) = \int_{y_1}^{y_1/p_\epsilon} \frac{2}{(1-p_\epsilon) y_2} dy_2 = \frac{2}{(1-p_\epsilon)} [ \ln(y_2) \big|_{y_1}^{y_1/p_\epsilon} ] = \frac{2\ln(p_\epsilon)}{(p_\epsilon - 1)}.
    \end{equation}
    For $p_\epsilon / 2 \leq y_1 \leq 1/2$ we have,
    \begin{equation}
        f_{y_1}(y_1) = \int_{y_1}^{1/2} \frac{2}{(1-p_\epsilon) y_2} dy_2 = \frac{2}{(1-p_\epsilon)} [ \ln(y_2) \big|_{y_1}^{1/2} ] = \frac{2\ln(2y_1)}{(p_\epsilon-1)}.
    \end{equation}
    Thus, as a piecewise function the PDF is formally,
    \begin{equation}
    \label{eq:uniform-pdf}
        f_{y_1}(y_1) = \begin{cases}
            \frac{2\ln(p_\epsilon)}{(p_\epsilon-1)} \quad &\text{ for } 0 \leq y_1 \leq \frac{p_\epsilon}{2},\\
            \frac{2\ln(2y_1)}{(p_\epsilon-1)} \quad &\text{ for } \frac{p_\epsilon}{2} \leq y_1 \leq 1/2.
        \end{cases}
    \end{equation}
    Now, the CDF is determined through integration,
    \begin{equation}
    \label{eq:uniform-cdf}
        F_{y_1}(y_1) = \int_0^{y_1} f_{y_1}(y_1)dy_1 =  \begin{cases}
            \frac{2y_1\ln(p_\epsilon)}{(p_\epsilon-1)} \quad &\text{ for } 0 \leq y_1 \leq \frac{p_\epsilon}{2},\\
            \frac{2y_1(\ln(2y_1) - 1)+ p_\epsilon}{(p_\epsilon-1)} \quad &\text{ for } \frac{p_\epsilon}{2} \leq y_1 \leq 1/2.
        \end{cases}
    \end{equation}
    We can integrate the CDF to get,
    \begin{equation}
        \int_{0}^{y_1} F_{y_1}(y_1) =  \begin{cases}
            \frac{y_1^2\ln(p_\epsilon)}{(p_\epsilon-1)} \quad &\text{ for } 0 \leq y_1 \leq \frac{p_\epsilon}{2},\\
            \frac{4y_1^2(2\ln(2y_1) - 3) + 8 y_1 p_\epsilon - p_\epsilon^2}{8(p_\epsilon-1)} \quad &\text{ for } \frac{p_\epsilon}{2} \leq y_1 \leq 1/2.
        \end{cases}
    \end{equation}
    Plugging all of this back into Equation~\ref{eq:bidding-strategy} yields,
    \begin{align}
        \hat{b}^*_i &= \begin{cases}
            p_\epsilon + v_i^p\frac{2v_i^p\ln(p_\epsilon)}{p_\epsilon-1} - \frac{(v_i^p)^2\ln(p_\epsilon)}{p_\epsilon-1},\\
            p_\epsilon + v_i^p\frac{2v_i^p(\ln(2v_i^p) - 1)+ p_\epsilon}{(p_\epsilon-1)} - \frac{4(v_i^p)^2(2\ln(2v_i^p) - 3) + 8 v_i^p p_\epsilon - p_\epsilon^2}{8(p_\epsilon-1)},
            \end{cases} \nonumber \\
        &= \begin{cases}
            p_\epsilon + \frac{(v_i^p)^2\ln(p_\epsilon)}{p_\epsilon-1} &\text{ if } 0 \leq v_i^p \leq \frac{p_\epsilon}{2},\\
            p_\epsilon + \frac{8(v_i^p)^2(\ln(2v_i^p) - 1/2)+ p_\epsilon^2}{8(p_\epsilon-1)} &\text{ if } \frac{p_\epsilon}{2} \leq v_i^p \leq \frac{1}{2}.
            \end{cases}
    \end{align}
    Since $b_i$ cannot be larger than 1, we cap the bidding function at one via,
    \begin{align}
        b_i^* :=  \min\{\hat{b}^*_i, 1\}.
    \end{align}
    The utility gained by agent $i$ for using such a bidding function is,
    \begin{equation}
        u(b_i^*) = \begin{cases}
            v_i^d - b_i^* +  \frac{2(v_i^p)^2\ln(p_\epsilon)}{p_\epsilon-1} \; &\text{ for } 0 \leq v_i^p \leq \frac{p_\epsilon}{2},\\
            v_i^d - b_i^* + \frac{2(v_i^p)^2(\ln(2v_i^p) - 1)+ p_\epsilon}{(p_\epsilon-1)} \; &\text{ for } \frac{p_\epsilon}{2} \leq v_i^p \leq 1/2.
            \end{cases}
    \end{equation}
    When this utility is larger than 0, the agent will participate otherwise the agent will not submit a model to the regulator.
    Finally, we can find the optimal compliance level by using Assumption \ref{assumption:a2},
    \begin{align}
        s_i^* := M^{-1}\big(b_i^*\big).
    \end{align}
\end{proof}

\subsection{Proof of Corollary 2}
\begin{repcorollary}{cor:beta-valuations}[Restated]
    Under Assumption \ref{assumption:a2}, let agents have total value $V_i$ and scaling factor $\lambda_i$ stem from Beta ($\alpha, \beta = 2$) and Uniform distributions respectively, with $v_i^d = (1-\lambda_i)V_i$ and $v_i^p = \lambda_iV_i$.
    Denote the CDF of the Beta distribution on $[0,1]$ as $F_\beta(x) = 3x^2 - 2x^3$.
    Optimal bid and utility for agents participating in \prb{Circa} (Equation~\ref{eq:all-pay-auction}) are,
    $$
         b_i^* := \min \{ \hat{b}_i^*, 1\}, \quad \hat{b}_i^* = \begin{cases}
        p_\epsilon + \frac{3(v_i^p)^2(p_\epsilon^2 - 2p_\epsilon + 1)}{1 - F_\beta(p_\epsilon)}  &\text{ if } 0 \leq v_i^p \leq \frac{p_\epsilon}{2},\\
            p_\epsilon + \frac{8(v_i^p)^2\big(6(v_i^p)^2 - 8v_i^p + 3 \big) + p_\epsilon^3(3p_\epsilon - 4)}{8(1 - F_\beta(p_\epsilon))} &\text{ if } \frac{p_\epsilon}{2} \leq v_i^p \leq \frac{1}{2},
        \end{cases}
    $$
    $$
        u(b_i^*; \; \bm{b}_{-i}, v_i^d, v_i^p) = \begin{cases}
            v_i^d +  \frac{6(v_i^p)^2(p_\epsilon^2 - 2p_\epsilon + 1)}{1 - F_\beta(p_\epsilon)} - b_i^* \; &\text{ for } 0 \leq v_i^p \leq \frac{p_\epsilon}{2},\\
            v_i^d + \frac{v_i^p\big(8(v_i^p)^3 - 12(v_i^p)^2 + 6v_i^p + p_\epsilon^2(2p_\epsilon - 3)\big)}{1 - F_\beta(p_\epsilon)} - b_i^* \; &\text{ for } \frac{p_\epsilon}{2} \leq v_i^p \leq 1/2.
            \end{cases}
    $$
    Participating agents submit models with compliance,
    $$
        s_i^* = \begin{cases}
            M^{-1}(b_i^*) > \epsilon &\text{ if } u_i(b_i^*; \; \bm{b}_{-i}, v_i^d, v_i^p) > 0,\\
            0 \text{ (no model submission) } &\text{ else}.
        \end{cases}
    $$
\end{repcorollary}

\begin{proof}

    Similar to Corollary \ref{cor:uniform-valuations}, we begin solving for $f_{v_i^p}(\cdot)$ using a change of variables.
    For the product of two random variables $v = x_1 \cdot x_2$, let $y_1 = x_1 \cdot x_2$ and $y_2 = x_2$.
    Inversely, $x_2 = y_2$ and $x_1 = y_1 / y_2$.
    While $x_1$ and $x_2$ are independent, $x_1$ comes from a Beta distribution and $x_2$ from a Uniform one.
    The PDF and CDF of a Beta distribution, with $\alpha = \beta = 2$, on $[0,1]$ are defined as,
    \begin{align}
        f_\beta(x) := 6x(1-x), \\
        F_\beta(x) := 3x^2 - 2x^3.
    \end{align}
    Now, the PDF over $y_1,y_2$ is defined as,
    \begin{equation}
        f_{y_1, y_2}(x_1, x_2) = (\frac{6x_1(1-x_1)}{1-F_\beta(p_\epsilon)})(\frac{1}{1/2-0}) = \frac{12x_1(1-x_1)}{1-F_\beta(p_\epsilon)}.
    \end{equation}
    When using the change of variables this becomes,
    \begin{equation}
        f_{y_1, y_2}(y_1, y_2) = f_{y_1, y_2}(x_1, x_2) |J|= \frac{12y_1(1-\frac{y_1}{y_2})}{(1-F_\beta(p_\epsilon))y_2^2}, \quad |J| = \bigg| \begin{pmatrix}
            1/y_2 & -y_1/y_2^2 \\ 0 & 1
        \end{pmatrix}\bigg| = 1/y_2
    \end{equation}
    Marginalizing out $y_2$ (a non-negative value) yields,
    \begin{equation}
        f_{y_1}(y_1) = \frac{12y_1}{1-F_\beta(p_\epsilon)} \int_{0}^{\infty} \frac{1}{y_2^2} - \frac{y_1}{y_2^3} dy_2.
    \end{equation}
    The bounds of integration depend upon the value of $y_1$.
    The change of variable to the $(y_1, y_2)$ space, where $0 \leq y_1, y_2 \leq 1/2$, results in a new region of possible variable values.
    This region is a triangle bounded by the three vertices: $(0,0)$, $(p_\epsilon/2, 1/2)$, and $(1/2, 1/2)$.
    Thus, the bounds of marginalization depend upon the value of $y_1$. 
    For $0 \leq y_1 \leq p_\epsilon/2$ we have,
    \begin{align}
        f_{y_1}(y_1) &= \frac{12y_1}{1-F_\beta(p_\epsilon)} \int_{y_1}^{y_1/p_\epsilon} \frac{1}{y_2^2} - \frac{y_1}{y_2^3} dy_2 = \frac{12y_1}{1-F_\beta(p_\epsilon)} [ -\frac{1}{y_2} + \frac{y_1}{2y_2^2} \big|_{y_1}^{y_1/p_\epsilon}]\nonumber\\
        &= \frac{12y_1}{1-F_\beta(p_\epsilon)}[-\frac{p_\epsilon}{y_1} + \frac{p_\epsilon^2}{2y_1} + \frac{1}{y_1} - \frac{1}{2y_1}] = \frac{6(p_\epsilon^2 - 2p_\epsilon + 1)}{1 - F_\beta(p_\epsilon)}.
    \end{align}
    For $p_\epsilon / 2 \leq y_1 \leq 1/2$ we have,
    \begin{align}
        f_{y_1}(y_1) &= \frac{12y_1}{1-F_\beta(p_\epsilon)} \int_{y_1}^{1/2} \frac{1}{y_2^2} - \frac{y_1}{y_2^3} dy_2 = \frac{12y_1}{1-F_\beta(p_\epsilon)} [ -\frac{1}{y_2} + \frac{y_1}{2y_2^2} \big|_{y_1}^{1/2}]\nonumber\\
        &= \frac{12y_1}{1-F_\beta(p_\epsilon)}[-2 + 2y_1 + \frac{1}{y_1} - \frac{1}{2y_1}] = \frac{6(4y_1^2 - 4y_1 + 1)}{1 - F_\beta(p_\epsilon)}.
    \end{align}
    Thus, as a piecewise function the PDF is formally,
    \begin{equation}
    \label{eq:beta-pdf}
        f_{y_1}(y_1) = \begin{cases}
            \frac{6(p_\epsilon^2 - 2p_\epsilon + 1)}{1 - F_\beta(p_\epsilon)} \quad &\text{ for } 0 \leq y_1 \leq \frac{p_\epsilon}{2},\\
            \frac{6(4y_1^2 - 4y_1 + 1)}{1 - F_\beta(p_\epsilon)} \quad &\text{ for } \frac{p_\epsilon}{2} \leq y_1 \leq 1/2.
        \end{cases}
    \end{equation}
    Now, the CDF is determined through integration,
    \begin{equation}
    \label{eq:beta-cdf}
        F_{y_1}(y_1) = \int_0^{y_1} f_{y_1}(y_1)dy_1 =  \begin{cases}
            \frac{6y_1(p_\epsilon^2 - 2p_\epsilon + 1)}{1 - F_\beta(p_\epsilon)} \quad &\text{ for } 0 \leq y_1 \leq \frac{p_\epsilon}{2},\\
            \frac{2y_1(4y_1^2 - 6y_1 + 3) + p_\epsilon^2(2p_\epsilon - 3)}{1 - F_\beta(p_\epsilon)} \quad &\text{ for } \frac{p_\epsilon}{2} \leq y_1 \leq 1/2.
        \end{cases}
    \end{equation}
    We can integrate the CDF to get,
    \begin{equation}
        \int_{0}^{y_1} F_{y_1}(y_1) =  \begin{cases}
            \frac{3y_1^2(p_\epsilon^2 - 2p_\epsilon + 1)}{1 - F_\beta(p_\epsilon)} \quad &\text{ for } 0 \leq y_1 \leq \frac{p_\epsilon}{2},\\
            \frac{8y_1\big(2y_1^3 - 4y_1^2 + 3y_1 + p_\epsilon^2(2p_\epsilon - 3)\big) + p_\epsilon^3(4 - 3p_\epsilon)}{8(1 - F_\beta(p_\epsilon))} \quad &\text{ for } \frac{p_\epsilon}{2} \leq y_1 \leq 1/2.
        \end{cases}
    \end{equation}
    Plugging all of this back into Equation~\ref{eq:bidding-strategy} yields,
    \begin{align}
        \hat{b}^*_i &= \begin{cases}
            p_\epsilon + v_i^p\frac{6v_i^p(p_\epsilon^2 - 2p_\epsilon + 1)}{1 - F_\beta(p_\epsilon)} - \frac{3(v_i^p)^2(p_\epsilon^2 - 2p_\epsilon + 1)}{1 - F_\beta(p_\epsilon)},\\
            p_\epsilon + v_i^p\frac{2v_i^p(4(v_i^p)^2 - 6v_i^p + 3) + p_\epsilon^2(2p_\epsilon - 3)}{1 - F_\beta(p_\epsilon)} - \frac{8v_i^p\big(2(v_i^p)^3 - 4(v_i^p)^2 + 3v_i^p + p_\epsilon^2(2p_\epsilon - 3)\big) + p_\epsilon^3(4 - 3p_\epsilon)}{8(1 - F_\beta(p_\epsilon))},
            \end{cases} \nonumber \\
        &= \begin{cases}
        p_\epsilon + \frac{3(v_i^p)^2(p_\epsilon^2 - 2p_\epsilon + 1)}{1 - F_\beta(p_\epsilon)}  &\text{ if } 0 \leq v_i^p \leq \frac{p_\epsilon}{2},\\
            p_\epsilon + \frac{8(v_i^p)^2\big(6(v_i^p)^2 - 8v_i^p + 3 \big) + p_\epsilon^3(3p_\epsilon - 4)}{8(1 - F_\beta(p_\epsilon))} &\text{ if } \frac{p_\epsilon}{2} \leq v_i^p \leq \frac{1}{2}.
        \end{cases}
    \end{align}
    Since $b_i$ cannot be larger than 1, we cap the bidding function at one via,
    \begin{align}
        b_i^* :=  \min\{\hat{b}^*_i, 1\}.
    \end{align}
    The utility gained by agent $i$ for using such a bidding function is,
    \begin{equation}
        u(b_i^*) = \begin{cases}
            v_i^d - b_i^* +  \frac{6(v_i^p)^2(p_\epsilon^2 - 2p_\epsilon + 1)}{1 - F_\beta(p_\epsilon)} \; &\text{ for } 0 \leq v_i^p \leq \frac{p_\epsilon}{2},\\
            v_i^d - b_i^* + \frac{v_i^p\big(8(v_i^p)^3 - 12(v_i^p)^2 + 6v_i^p + p_\epsilon^2(2p_\epsilon - 3)\big)}{1 - F_\beta(p_\epsilon)} \; &\text{ for } \frac{p_\epsilon}{2} \leq v_i^p \leq 1/2.
            \end{cases}
    \end{equation}
    When this utility is larger than 0, the agent will participate otherwise the agent will not submit a model to the regulator.
    Finally, we can find the optimal compliance level by using Assumption \ref{assumption:a2},
    \begin{align}
        s_i^* := M^{-1}\big(b_i^*\big).
    \end{align}

\end{proof}

\section{Additional Experiments}
\label{app:experiments}

\begin{figure}[!htbp]
    \centering
    \subfigure{
    \includegraphics[width=0.425\textwidth]{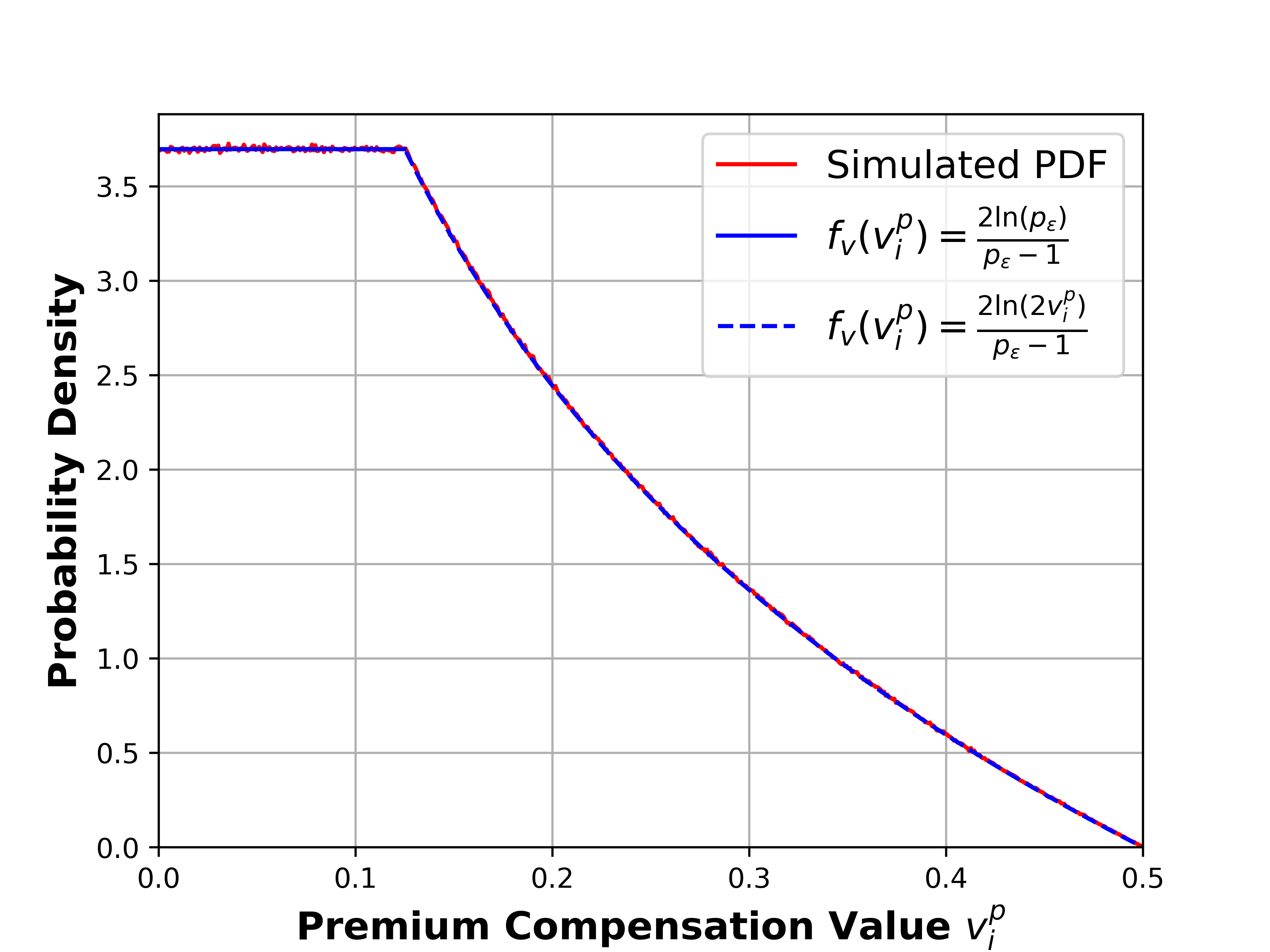}
    }
    \subfigure{
    \includegraphics[width=0.425\textwidth]{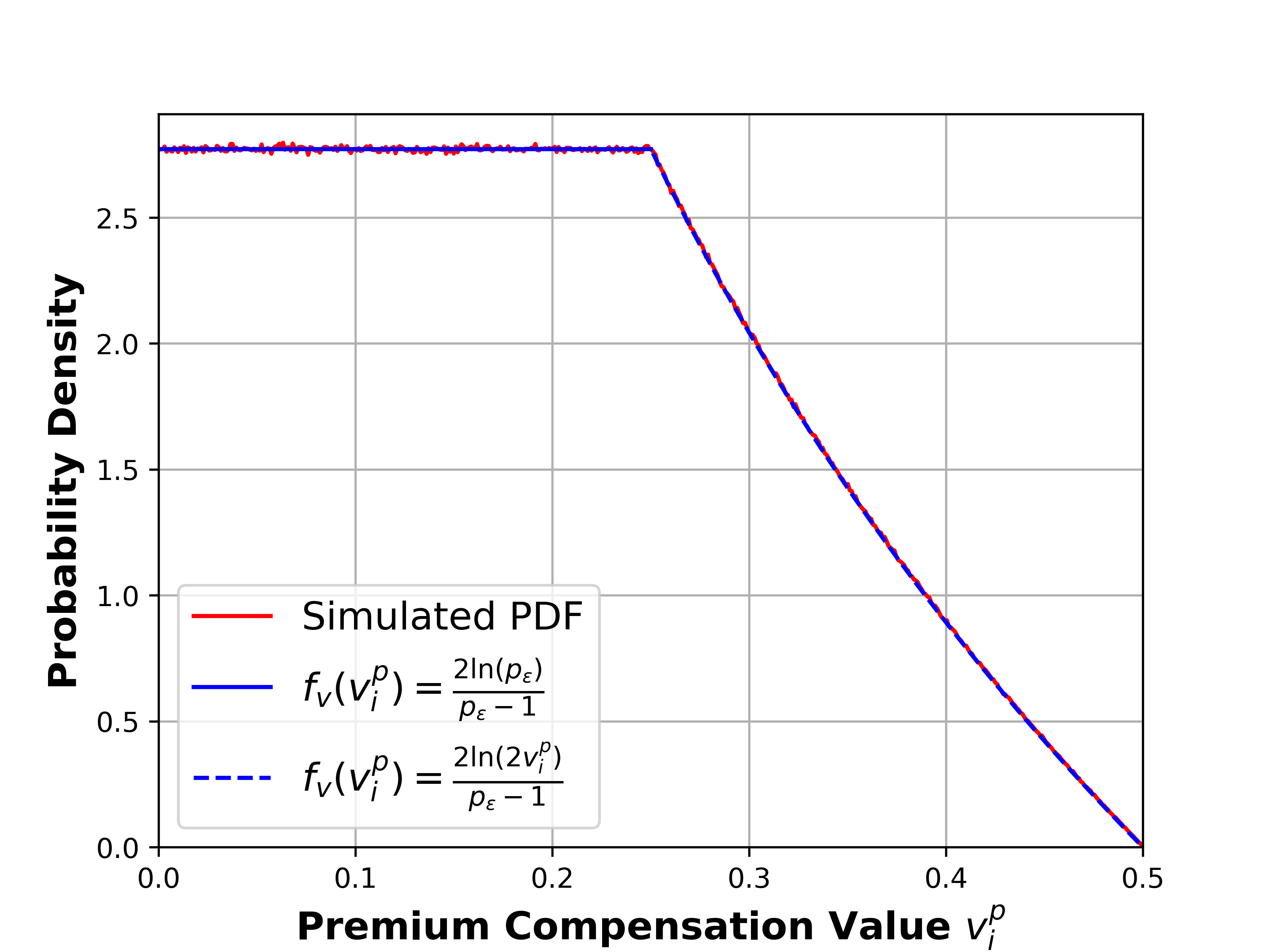}
    }
    \subfigure{
    \includegraphics[width=0.425\textwidth]{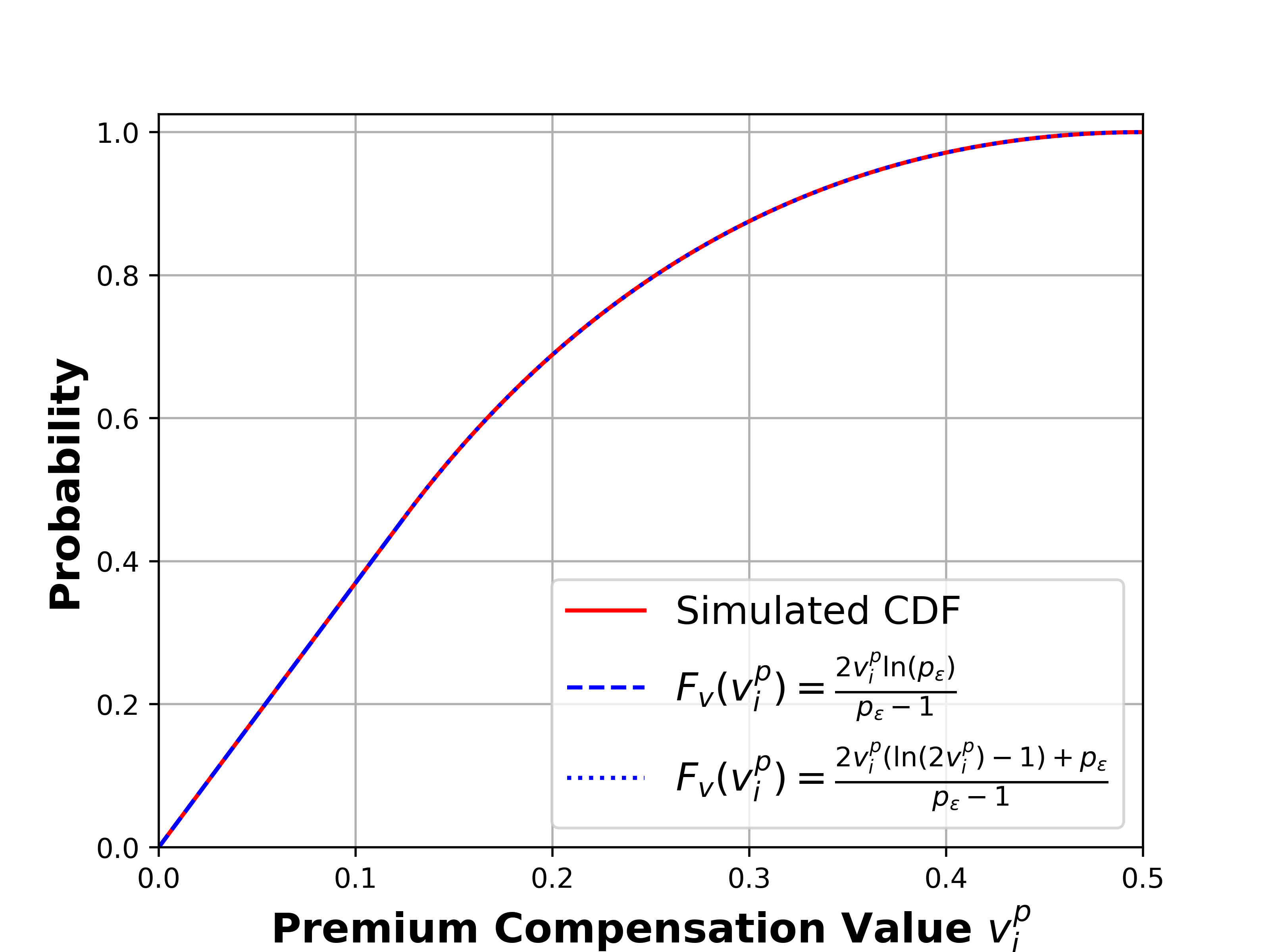}
    }
    \subfigure{
    \includegraphics[width=0.425\textwidth]{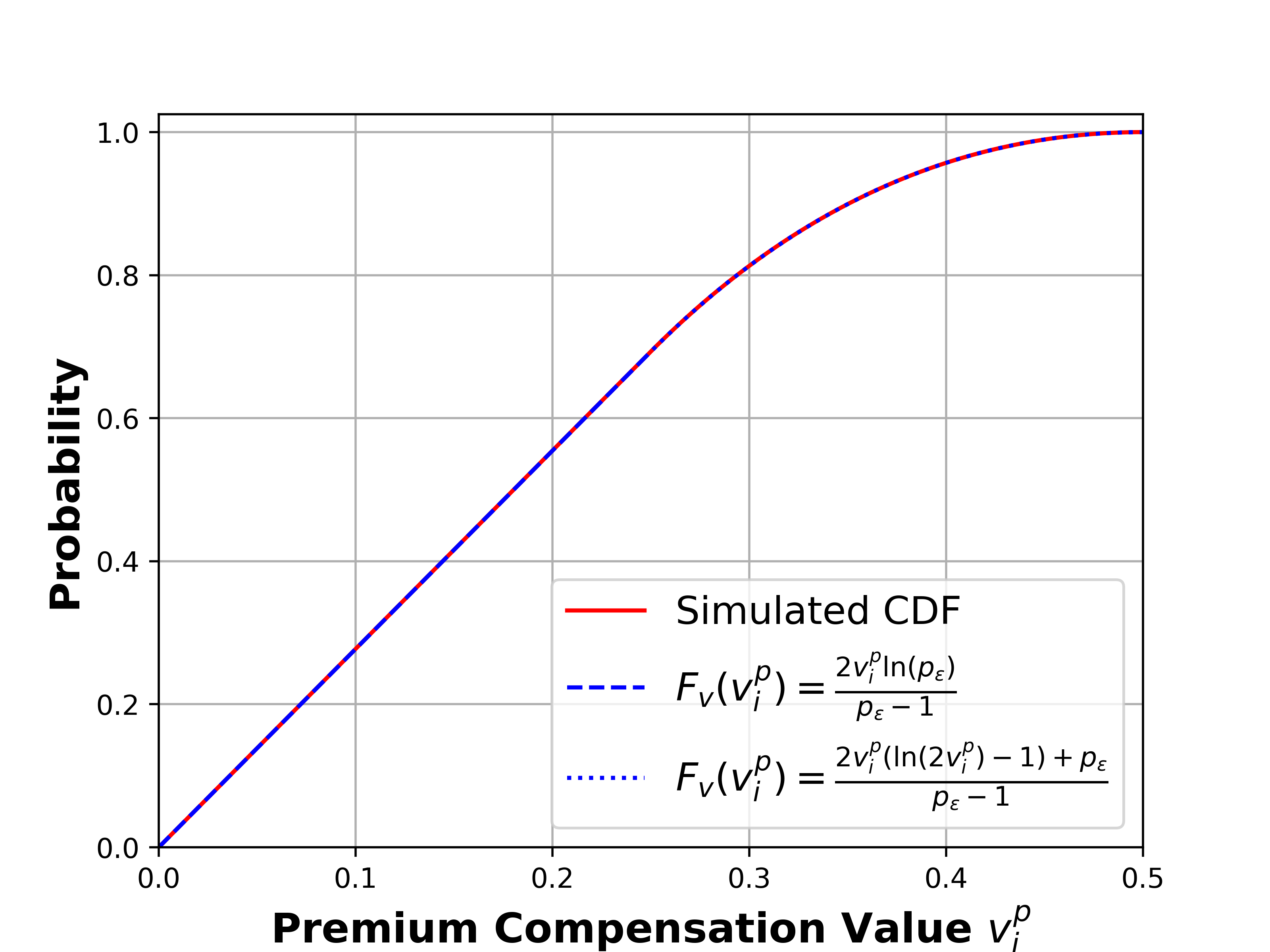}
    }
    \caption{Numerical validation of our derivations for $f_{v}(v_i^p)$ and $F_{v}(v_i^p)$, where $v_i^p := V_i\lambda_i$, for $V_i$ and $\lambda_i$ coming from Uniform distributions (Corollary \ref{cor:uniform-valuations}). The price of attaining $\epsilon$ is set as $p_\epsilon = 1/4$ (top row) and $p_\epsilon = 1/2$ (bottom row).}
    \label{fig:uniform-pdf-cdf}
\end{figure}

\begin{figure}[!htbp]
    \centering
    \subfigure{
    \includegraphics[width=0.425\textwidth]{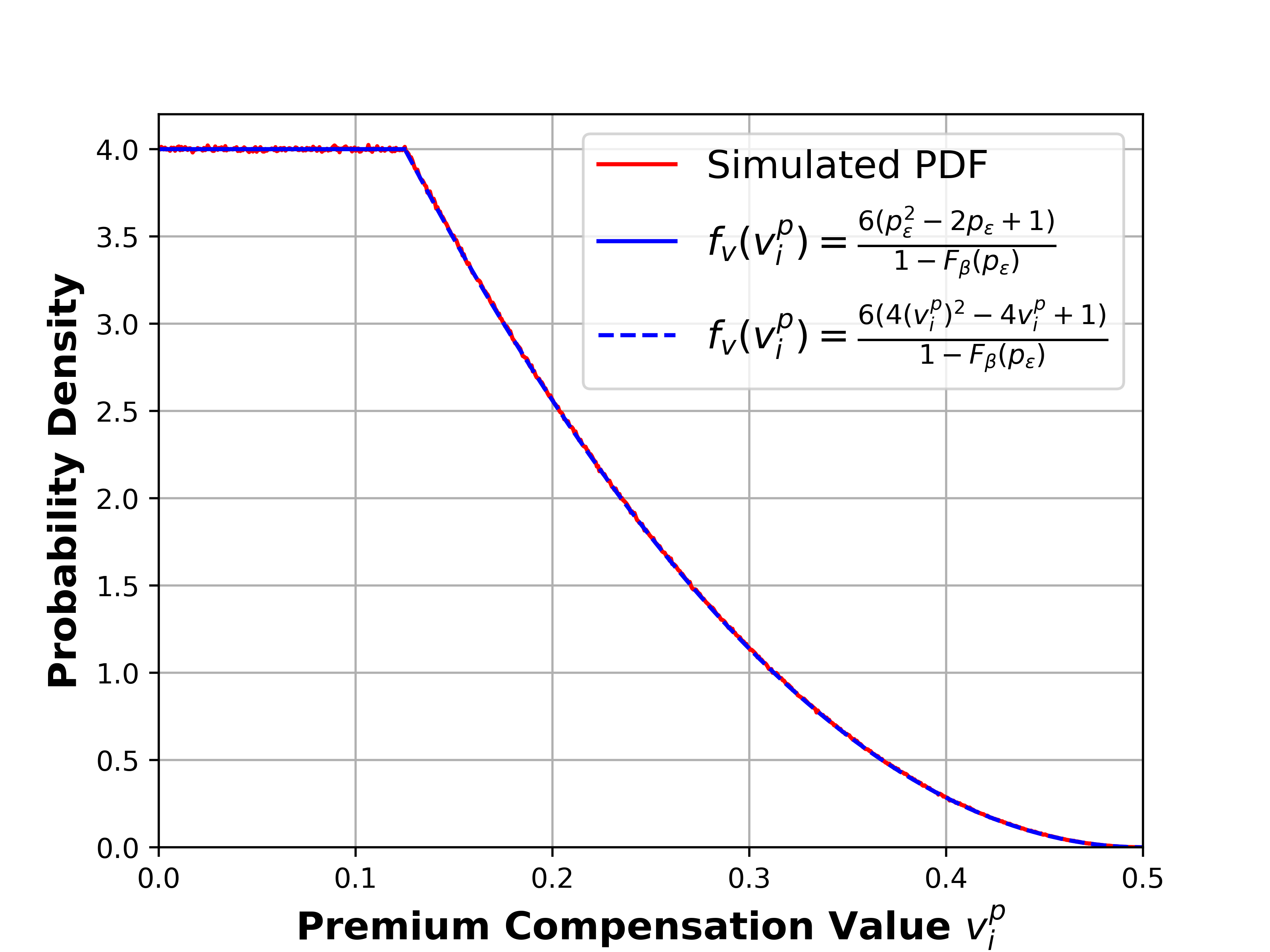}
    }
    \subfigure{
    \includegraphics[width=0.425\textwidth]{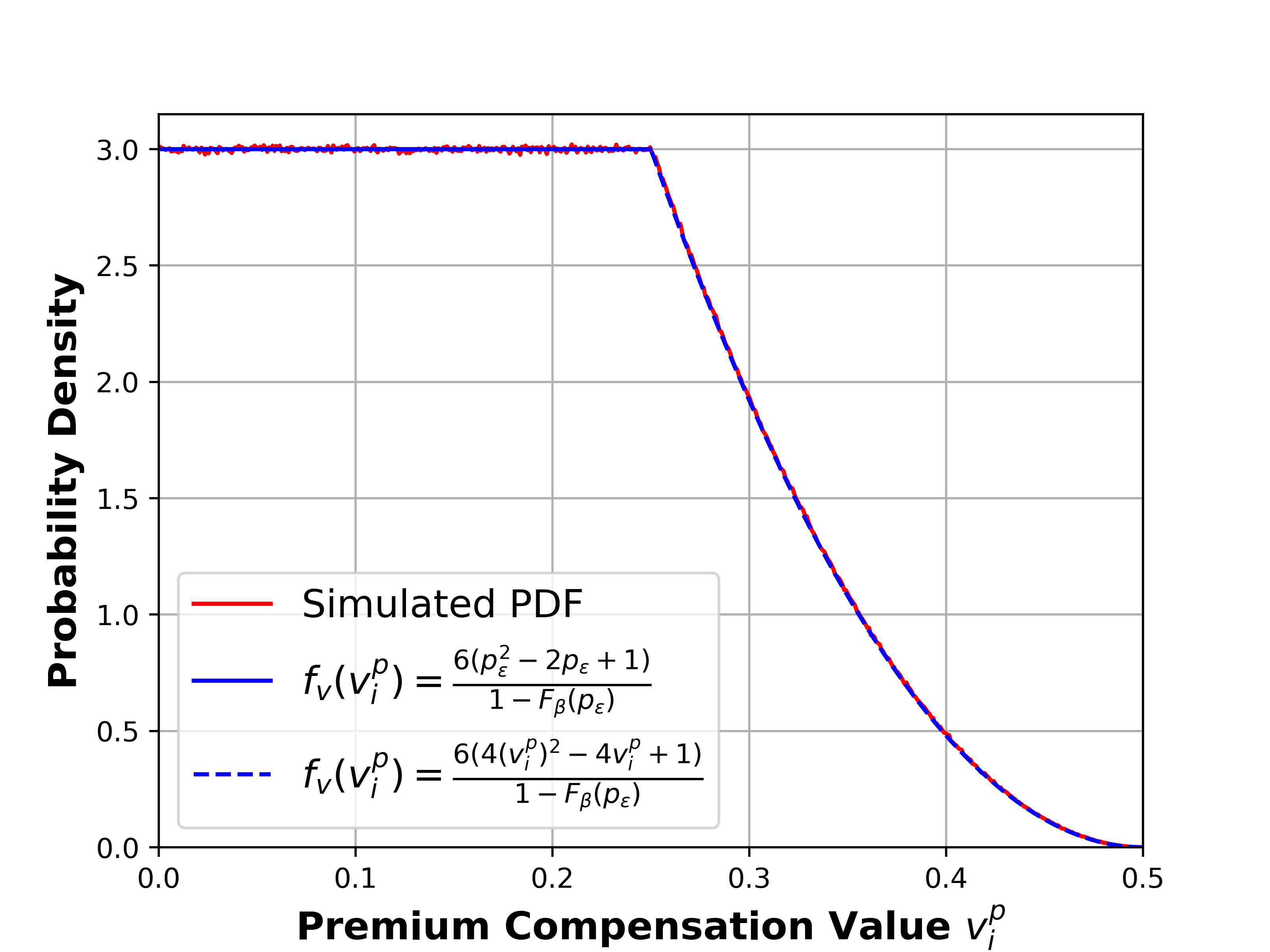}
    }
    \subfigure{
    \includegraphics[width=0.425\textwidth]{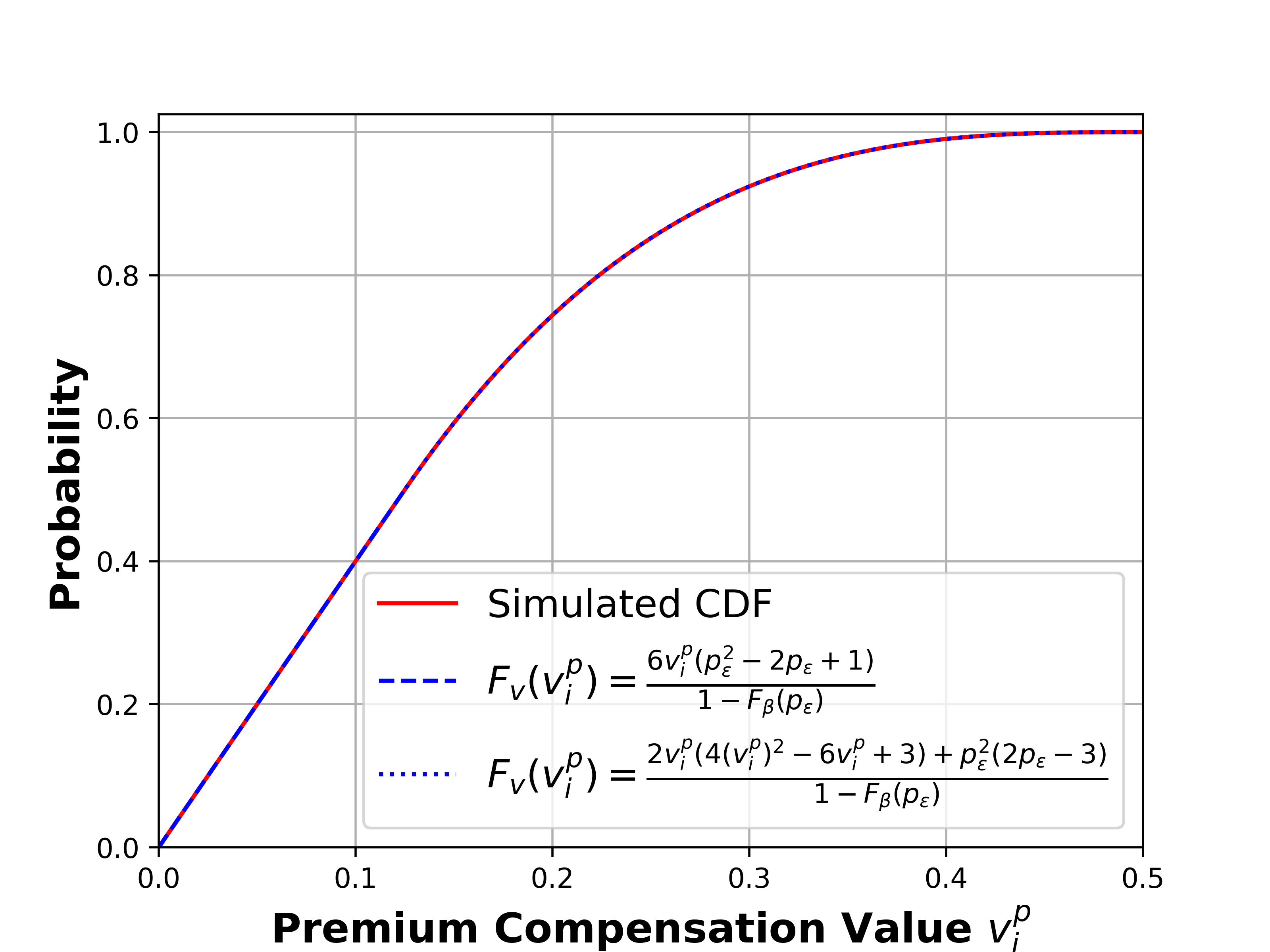}
    }
    \subfigure{
    \includegraphics[width=0.425\textwidth]{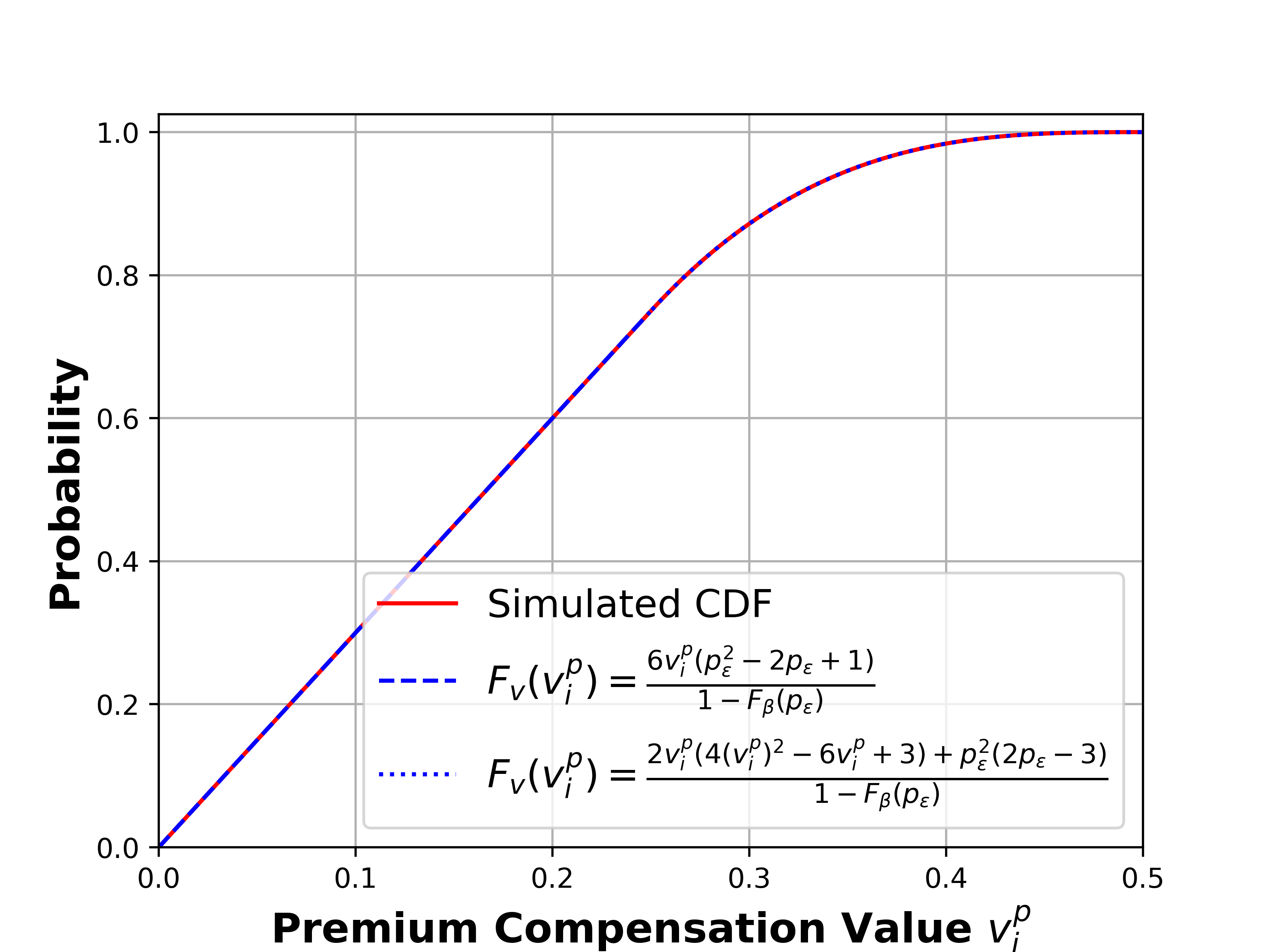}
    }
    \caption{Numerical validation of our derivations for $f_{v}(v_i^P)$ and $F_{v}(v_i^P)$, where $v_i^p := V_i\lambda_i$, for $V_i$ coming from a Beta distribution and $\lambda_i$ from a Uniform distributions (Corollary \ref{cor:beta-valuations}). The price of attaining $\epsilon$ is set as $p_\epsilon = 1/4$ (top row) and $p_\epsilon = 1/2$ (bottom row).}
    \label{fig:beta-pdf-cdf}
\end{figure}

Within this section, we verify empirically that our computed PDF and CDFs in Corollaries \ref{cor:uniform-valuations} and \ref{cor:beta-valuations} are correct.
To accomplish this, we randomly sample and compute the product of $V_i$ and $\lambda_i$ fifty million times.
We then plot the PDF and CDF of the resultant products and compare it with our theoretical PDF and CDF.
The theoretical PDF and CDF for Corollary \ref{cor:uniform-valuations} are defined in Equations~\ref{eq:uniform-pdf} and \ref{eq:uniform-cdf}, while those for Corollary \ref{cor:beta-valuations} are found in Equations~\ref{eq:beta-pdf} and \ref{eq:beta-cdf}.
The results of these simulations, which validate our computed PDFs and CDFs, are shown in Figures \ref{fig:uniform-pdf-cdf} and \ref{fig:beta-pdf-cdf}.
To ensure correctness, we perform testing on different values of $p_\epsilon$.
As expected, our theory lines up exactly with our empirical simulations for both Corollaries as well as across varying $p_\epsilon$.
We note that all experiments are computationally light, with all run locally on an M3 chip with 16GB of RAM. 

\section{Binary and Discrete Compliance in Circa}

Our framework still works within binary or discrete settings. 
This is important when dealing with properties or metrics that are not continuous, like how the EU AI Act evaluates AI risk into minimal, limited, high, and unacceptable tiers \citep{act2024eu}.
The rationale behind why \textsc{Circa} works for binary or discrete settings is that models can still be ranked or compared against each other depending on how well they satisfy the given metric or property.

For example, models can be separated into Pass/Fail categories, where the Pass category can be further split into High/Medium/Low sub-categories. 
All models achieving at least Low Pass compliance are cleared for deployment.
While a model either complies or does not, the models can still be gauged on how well they comply (e.g., High/Medium/Low).
Since a ranking of models can still be generated, premium rewards can be provided to higher-passing models.

In situations where the regulatory policy is black and white, for example \enquote{your model must be trained with differential privacy}, \textsc{Circa} still holds as an ordering or ranking between models can still be ascertained. 
In the example of differential privacy, \textit{any} model that is trained with differential privacy would be cleared for deployment.
However, it is also true that differential privacy can be gauged by a level of privacy $\epsilon_{DP}$ (not to be confused with our compliance threshold $\epsilon$). 
Models with smaller values of $\epsilon_{DP}$ will be provided additional premium rewards since they are more compliant (\textit{i.e.,} more private). 
Thus, \textsc{Circa} would still incentivize agents to become more private even when there is a binary compliance metric.

\section{Repeating \textsc{Circa} Auctions}
\label{app:repeating-auctions-full}

The current auction structure (Algorithm \ref{alg:circa}) expects agents to submit a single model trained solely for the upcoming auction. 
There is no expectation that the model will be reused for a future auction, or indication that the model has been submitted to a previous auction. 
Looking towards the future, we would like to design \textsc{Circa} to fit a repeatable auction structure, in which approved or rejected models may be resubmitted in subsequent auctions.

\textbf{Repeated Agent Utility}.
Previously, in Algorithm \ref{alg:circa}, 
agents start the regulatory process with zero cost and value (\textit{i.e.,} they are building their models from scratch).
In repeating \textsc{Circa} auctions, agent cost and value are accumulated across all previous auction submissions.
For example, if an agent trains its already-accepted model further to attain a higher compliance level $s_i$, its total accumulated training cost is $M(s_i)$.
This agent's total value becomes the value its model gained from previous auction submissions plus any value gained from the current auction.

By allowing repeated \textsc{Circa} auctions, an agent is able to repeatedly submit its model for regulatory review. We note that repeated submissions decrease the value of model deployment; once an agent earns the reward for deploying their model, subsequent deployments of the same model with improved compliance levels can be realistically expected to earn less value than the initial deployment.
We characterize this loss in value for repeated submissions with an indicator function in the utility function that only allows deployment value to be obtained once, on initial acceptance of a model. 
While we allow agents to win premium rewards across multiple auctions, we note that a regulator can curb this by either limiting the number of auction submissions per agent or the number of auctions held per year.
We now define the repeated \textsc{Circa} auction utility of agent $i$, who has participated in $a - 1$ previous auctions, as:
\begin{equation}
    u_{i, a}(b_i) = \left(\sum_{n=1}^{a} \nu_{i}^n\right) - b_i,
\end{equation}
where $\nu_i^n$, the value gained at the $n^{th}$ auction model $i$ was submitted to, is formulated as:
\begin{equation}
    \label{eq:all-pay-repeat-auction}
    \nu_i^n = \begin{cases}
        v_i^{d,n}\ \cdot 1_{\text{(if $\nu^{n-1}_i = 0$)}} & \text{if $b_j^n \geq p_\epsilon^n$ and } b_i^n < b_j^n \text{ randomly sampled bid } b_j^n,\\
        v_i^{d,n} \cdot 1_{\text{(if $\nu^{n-1}_i = 0$)}} + v_i^{p,n} & \text{if $b_i^n \geq p_\epsilon^n$ and } b_i^n > b_j^n \text{ randomly sampled bid } b_j^n,\\
        0 & \text{if n $\leq$ 0}.
    \end{cases}
\end{equation}
The repeated \textsc{Circa} auction setup creates a unique property for models in training. 
If an agent intends to obtain a high compliance level, but an auction takes place mid-training, the agent is actually incentivized to submit their model early if they have a chance at winning the premium reward. 
Though the model may have a lower likelihood of earning the reward, there is no consequence for models failing to attain the premium reward. 
Gaining value is strictly beneficial to agents, and accumulated value helps offset the costs of training a model. 
This property only exists for the premium reward; the deployment reward can only be obtained once, thus there is no incentive to submit early to earn it.

\textbf{Repeated Optimal Bidding Function}.
Using the same assumptions for single-auction \textsc{Circa}, namely Assumptions \ref{assumption:a1} and \ref{assumption:a2} along with private values, we can derive the bidding function for a rational agent under a repeated \textsc{Circa} auction setting.
We follow an equivalent setup to Theorem \ref{thm:general-bidding-strategy} with regards to the valuation of rewards, giving us the cumulative distribution function for $v^p_i = V_i\lambda_i$ as $F_{v}(\cdot)$ and the probability distribution function as $f_{v}(\cdot)$.

From our definition of utility $u_{i,a}(b_i)$, we find that an agent $i$ that does not participate (\textit{i.e.,} submitting $b_i = 0$) receives utility equal to $\nu^a_i$. 
However, since $b_i = 0$ will never be larger than $p_\epsilon$ (by definition), it must be true that $\nu^a_i = 0$ as well, since the model will never meet the required compliance threshold.
Therefore, a non-participating agent will always receive zero utility.
\begin{equation}
    \label{eq:zero-bid-repeating}
    u_{i,a}(0) = 0.
\end{equation}
Following a similar proof structure as Theorem \ref{thm:general-bidding-strategy} in Appendix \ref{app:proofs}, we find that participating agents $i \in P$ (with $P$ defined in the previous proof) will now have a utility of,

\begin{align}
    u_{i,a}(b_i) &= \nu^a_i + v^d_i \cdot 1_{\text{($\nu^{a}_i$ = 0)}} + v_i^p \mathbb{P}\big(b_i > b_j \big) - b(b_i), \quad b_j \sim \text{randomly sampled agent bid}, \nonumber\\
     &= \nu^a_i + v^d_i \cdot 1_{\text{($\nu^{a}_i$ = 0)}} + v_i^p F_{v}(b_i) - b(b_i).
\end{align}
Taking the derivative and setting it equal to zero yields,
\begin{equation}
    \frac{d}{d b_i} u_{i,a}(b_i) = v_i^p f_{v}(b_i) - b'(b_i) = 0.
\end{equation}
As agents bid in proportion to their valuation, we solve the first-order conditions at $b_i = v_i^p$,
\begin{equation}
    b'(v_i^p) = v_i^p f_{v}(v_i^p).
\end{equation}
Note, at this point in the proof the bidding function calculation is now equivalent to the calculations found in Theorem \ref{thm:general-bidding-strategy}. We can thus follow the same steps to reveal our optimal bidding function,
\begin{align}
    b(v_i^p) :&= p_\epsilon + v_i^p F_{v}(v_i^p) - \int_0^{v_i^p} F_{v}(z)dz,
\end{align}
which is equivalent to the optimal bidding function derived in Theorem \ref{thm:general-bidding-strategy}.

As the optimal bidding function is equivalent, calculations for the Nash Bidding Equilibrium are also equivalent to those found in Corollary \ref{cor:uniform-valuations} and Corollary \ref{cor:beta-valuations}. The optimal bid and utility participating in \textsc{Circa} Equation~\ref{eq:all-pay-auction} under the assumptions of Corollary \ref{cor:uniform-valuations} will thus be,
$$
    b_i^* := \min \{ \hat{b}_i^*, 1\}, \quad \hat{b}_i^* = \begin{cases}
        p_\epsilon + \frac{(v_i^p)^2\ln(p_\epsilon)}{p_\epsilon-1} &\text{ if } 0 \leq v_i^p \leq \frac{p_\epsilon}{2},\\
        p_\epsilon + \frac{8(v_i^p)^2(\ln(2v_i^p) - 1/2)+ p_\epsilon^2}{8(p_\epsilon-1)} &\text{ if } \frac{p_\epsilon}{2} \leq v_i^p \leq \frac{1}{2},
    \end{cases}
$$
$$
    u_{i,a}(b_i^*; \bm{b}_{-i}) = 
    \begin{cases}
    \nu^a_i + v^d_i \cdot 1_{\text{($\nu^{a}_i$ = 0)}} + \frac{2(v_i^p)^2\ln(p_\epsilon)}{p_\epsilon-1} -  b_i^* &\text{ if } 0 \leq v_i^p \leq \frac{p_\epsilon}{2}, \\
    \nu^a_i + v^d_i \cdot 1_{\text{($\nu^{a}_i$ = 0)}} + \frac{2(v_i^p)^2(\ln(2v_i^p)-1) + p_\epsilon}{p_\epsilon-1} -  b_i^* &\text{ if } \frac{p_\epsilon}{2} \leq v_i^p \leq \frac{1}{2}.
    \end{cases}
$$
Agents participating in \textsc{Circa} under Corollary \ref{cor:uniform-valuations} submit models with the following compliance,
$$
    s_i^* := \begin{cases}
         M^{-1}(b_i^*) > \epsilon &\text{ if } u_i(b_i^*; \bm{b}_{-i}) > 0,\\
        0 \text{ (no model submission) } &\text{ else}.
    \end{cases}
$$

The optimal bid and utility participating in \textsc{Circa} Equation~\ref{eq:all-pay-auction} under the assumptions of Corollary \ref{cor:beta-valuations} will be,
$$
     b_i^* := \min \{ \hat{b}_i^*, 1\}, \quad \hat{b}_i^* = \begin{cases}
    p_\epsilon + \frac{3(v_i^p)^2(p_\epsilon^2 - 2p_\epsilon + 1)}{1 - F_\beta(p_\epsilon)}  &\text{ if } 0 \leq v_i^p \leq \frac{p_\epsilon}{2},\\
        p_\epsilon + \frac{8(v_i^p)^2\big(6(v_i^p)^2 - 8v_i^p + 3 \big) + p_\epsilon^3(3p_\epsilon - 4)}{8(1 - F_\beta(p_\epsilon))} &\text{ if } \frac{p_\epsilon}{2} \leq v_i^p \leq \frac{1}{2},
    \end{cases}
$$
$$
\resizebox{0.975\hsize}{!}{
    $u_{i,a}(b_i^*; \bm{b}_{-i}) = \begin{cases}
        \nu^a_i + v^d_i \cdot 1_{\text{($\nu^{a}_i$ = 0)}} +  \frac{6(v_i^p)^2(p_\epsilon^2 - 2p_\epsilon + 1)}{1 - F_\beta(p_\epsilon)} - b_i^* \; &\text{ for } 0 \leq v_i^p \leq \frac{p_\epsilon}{2},\\
        \nu^a_i + v^d_i \cdot 1_{\text{($\nu^{a}_i$ = 0)}} + \frac{v_i^p\big(8(v_i^p)^3 - 12(v_i^p)^2 + 6v_i^p + p_\epsilon^2(2p_\epsilon - 3)\big)}{1 - F_\beta(p_\epsilon)} - b_i^* \; &\text{ for } \frac{p_\epsilon}{2} \leq v_i^p \leq 1/2.
        \end{cases}$
        }
$$
Agents participating in \textsc{Circa} under Corollary \ref{cor:beta-valuations} submit models with the following compliance,
$$
    s_i^* = \begin{cases}
        M^{-1}(b_i^*) > \epsilon &\text{ if } u_i(b_i^*; \bm{b}_{-i}) > 0,\\
        0 \text{ (no model submission) } &\text{ else}.
    \end{cases}
$$

\section{Future Work}
\label{app:future-work}

While this work addresses key challenges in regulating AI compliance, several directions remain open for future exploration:

\textit{(1) Model Evaluation:} Creating a realistic protocol for the regulator to evaluate submitted model compliance levels is important to ensure agents do not skirt around compliance requirements. While we leave this problem for future work, one possible solution is that agents can either provide the regulator API access to test its model or provide the model weights directly to the regulator. Truthfulness can be enforced via audits and the threat of legal action.

\textit{(2) Extension to Heterogeneous Settings:} Extending our mechanism to heterogeneous scenarios, where evaluation data for agents and regulators differs, is a critical next step. 
Real-world data distributions often vary across contexts, and understanding how these variations affect both model compliance and agent strategies will create a more robust regulatory mechanism. 
While explicit protocols or mathematical formulations are left as future work, we have a few ideas.
One idea could be establishing a data-sharing framework between agents and the regulator, where each participating agent must contribute part of (or all of) its data to the regulator for evaluation.
If data can be anonymized, then this would be a suitable solution.
Another idea could be that the regulator collects data on its own, and can compare its distribution of data versus each participating agents' data distribution.
If distributions greatly differ, then the regulator could collect more data or resort to the previous data-sharing method.

\end{document}